\documentclass[useAMS,usenatbib]{mn2e}

\usepackage[a4paper, margin=2cm]{geometry}
\usepackage{pdflscape}

\usepackage{float,color}
\usepackage{epsf,rotating,url}
\usepackage{amssymb,amsfonts,amstext,amsgen,amsopn,amsxtra,indentfirst,lscape,times,rotating}
\usepackage{longtable,subfigure}
\usepackage{graphics}
\usepackage{keyval}
\usepackage{trig}
\usepackage[normalem]{ulem}
\usepackage{dcolumn}
\usepackage{bm}
\usepackage{graphicx}
\usepackage{txfonts}
\usepackage{multirow}

\newcommand\ion[2]{#1$\;${\scshape{#2}}}

\title[Variability in Active Galaxies]{Line and Continuum Variability in Active Galaxies}
  
\author[Y. E. Rashed et al.]{Y. E. Rashed,$^{1,2}$\thanks{yasir@ph1.uni-koeln.de}
  A. Eckart,$^{1,3}$ M. Valencia-S.,$^1$ M. Garc\'{i}a-Mar\'{i}n,$^1$
  G. Busch,$^1$ 
  \newauthor J. Zuther,$^1$ M. Horrobin,$^1$ and H. Zhou$^{4,5}$\\
  $^1$I. Physikalisches Institut, Universit\"at zu K\"oln, Z\"ulpicher Stra\ss e 77, 50937 K\"oln, Germany\\
  $^2$Department of Astronomy, Faculty of Science, University of Baghdad, 10071 Baghdad - Aljadirya, Iraq\\
  $^3$Max-Planck-Institut f\"ur Radioastronomie, Auf dem H\"ugel 69, 53121 Bonn, Germany\\
  $^4$Key laboratory for Research in Galaxies and Cosmology, Department of Astronomy,\\
      The University of Sciences and Technology of China, Chinese Academy of Sciences, Hefei, Anhui 230026, China\\
  $^5$Polar Research Institute of China, Jinqiao Rd. 451, Shanghai 200136, China; zhouhongyan@pric.gov.cn}

\begin{document}

\date{Accepted 2015 September 3.  Received 2015 August 24; in original form 2015 July 3}

\pagerange{\pageref{firstpage}--\pageref{lastpage}} \pubyear{0000}

\maketitle

\label{firstpage}

\begin{abstract}
We compared optical spectroscopic and photometric data for 18 AGN galaxies over 2 to 3 epochs, with
time intervals of typically 5 to 10 years. We used the Multi-Object Double Spectrograph (MODS) at the
Large Binocular Telescope (LBT) and compared the spectra to data taken from the SDSS database and the literature.
We find variations in the forbidden oxygen lines as well as in the hydrogen recombination lines of these sources.
For  4 of the sources we find that, within the calibration uncertainties, the variations in continuum and line spectra
of the sources are very small. 
We argue that it is mainly the difference in black hole mass 
between the samples that is responsible for the different degree of continuum variability.
In addition we find that for an otherwise constant accretion rate the total line variability 
(dominated by the narrow line contributions) reverberates the continuum variability with a dependency 
$\Delta L_{line} \propto (\Delta L_{cont.})^{\frac{3}{2}}$.
Since this dependency is prominently expressed in the narrow line emission it implies that the luminosity dominating part 
of the narrow line region must be very compact with a size of the order of at least 10 light years.
A comparison to literature data shows that these findings describe the variability 
characteristics of a total of 61 broad and narrow line sources.
\end{abstract}


\section{Introduction}
Seyfert galaxies and quasars are the main 
types of active galactic nuclei (AGNs) in which the nuclei are
dominated by a very powerful electromagnetic emitter. 
Seyfert nuclei are 10 to 100 times brighter than the host galaxies 
they reside in. Thus, these nuclei are responsible for most 
activity that occurs within the central 10-100~pc of these hosts.
 
Seyfert galaxies can show a wide variety of forbidden and permitted emission lines, which can have widths up to $\sim 10\ 000\,\mathrm{km}\,\mathrm{s}^{-1}$ via Doppler broadening. Depending on the width of the emission lines, one distinguishes between two main categories, Seyfert-1 (S1) and Seyfert-2 (S2):
S2 galaxies show only narrow lines ($\lesssim 1000\,\mathrm{km}\,\mathrm{s}^{-1}$), while S1 show additional broad components ($\lesssim 10\ 000\,\mathrm{km}\,\mathrm{s}^{-1}$) in the permitted lines. 
The emission lines are commonly believed to stem from the ``broad line region'' (BLR) and the 
``narrow line region'' (NLR), which are gas clouds orbiting a central super-massive black hole (SMBH) at very high velocities. The gas clouds are excited by continuum radiation produced by an accretion disk of material infalling around the central SMBH.
According to the unified model \citep[e.g.,][]{1993ARA&A..31..473A,1995PASP..107..803U}, the existence of S1 galaxies which have broad and narrow lines, and S2 galaxies which only have narrow lines, can be attributed to inclination effects. One possible scenario is a dusty torus which surrounds the accretion disk and BLR, obscuring the emission from the BLR if the galaxy is observed with a viewing angle parallel to the torus plane.

Narrow-line Seyfert-1 (NLS1) galaxies are S1 galaxies that have broad components with comparably narrow widths ($\lesssim 2000\,\mathrm{km}\,\mathrm{s}^{-1}$). From an observational point of view it appears that there are two main 
differences between S2 and NLS1 galaxies. The first one is that NLS1 objects 
have \ion{Fe}{ii} in their spectra (which originate in the BLR), which is 
absent in the S2 spectra \citep{1996A&A...314..419G}. The second  
is that some NLS1 sources show strong lines of highly ionized iron in their 
spectra e.g. [\ion{Fe}{vii}]$\lambda 5721$ \AA\ and [\ion{Fe}{x}]$\lambda 6375$ \AA\  
\citep{1985PASP...97..906O, 1985ApJ...297..166O}, these lines are rare and not 
typical for the S2 spectra.\\

AGNs are characterized by variability at almost all wavelengths \citep{2001sac..conf....3P}. 
Studying the variability on all time scales is a very useful approach to investigate and 
understand the physical properties of AGNs, 
there are many systematic studies on the variability of AGN using 
spectroscopic and  photometric observations. 
Photometric measurements have been used to study the variability, despite 
the fact that bright line emission is located within the corresponding passband \citep{2004ApJ...601..692V}. 
Studying the spectral variability both in line and in continuum emission has become an important topic 
that may help to outline differences between different Seyfert source classes. Variability may indicate 
how the active galactic nuclei (AGN) affect the regions surrounding them (i.e. the BLR and NLR) and how
these regions change as a function of time.  Many interesting results have been obtained from 
AGN samples of various types and sizes. A noted one is the anti-correlation between the continuum luminosity and variability, 
which was first reported by \cite{1972IAUS...44..171A}. Later additional information on this effect was provided 
by \cite{1994MNRAS.268..305H, 2012ApJ...758..104Z, 2013IAUS..290..373Z}. Additionally,
other authors \citep{2000AAS...196.5004C, 2013A&A...560A.104M} have clarified that 
this anti-correlation can qualitatively be described via the standard accretion disk model,
proposing that the variability is caused via the variation of the accretion rate.\\

Since spectroscopic studies can be very time consuming,
there are only a few detailed, spectroscopy-based comparative studies on the variability of AGNs.
They either concentrate on 
reverberation mapping analysis of extensively observed 
sources \citep[e.g.,][]{2007ApJ...659..997K} or depend on just a 
few epochs of spectroscopic observations \citep[e.g.,][]{2005ApJ...633..638W}.  
Multi epoch spectroscopy observations have advantages compared to photometric observations, as 
they allow us to study the continuum and line emission at the same time. 
Excepting regions of high emission line density, 
these observations also allow us to extract the underlying continuum spectrum 
over a broad wavelength range and to study 
the spectral shape of the lines and the line spectrum in general.
The emission line spectra then allow us to study the BLR and NLR regions surrounding 
the compact variable continuum nuclei.\\

Examples of extensive investigations on the variability in Seyferts and quasars concentrating on the 
continuum and the fluxes of the emission lines are, e.g., 
\cite{1984ApJ...279..529P, 1986AJ.....92..552P, 1998ApJ...501...82P, 2014ApJ...792...33G}. 
While the BLR lines are found to reverberate the continuum variability on time scales of days,
there are also a few reports of long term line variability in the NLR region:
Based on a time coverage of 8 years \citet{ClavelWamsteker1987} find for 3C390.3
that the NLR is photoionized and reverberates the continuum flux. They set an upper limit of 10 light years for the size of the NLR. 
Based on a time coverage of 5 years \citet{Peterson2013} find for the S1 galaxy NGC~5548 
that the [O III] $\lambda \lambda$4959, 5007 emission-line flux varies with time and give a size estimate of 1-3~pc (i.e. 3-9~light years).
There is a clear connection between the continuum and line variability. 
For instance, \cite{1984ApJ...279..529P} found that for $\sim$73\% of the cases they studied, there is variability 
in the $\mathrm{H\beta}$ line flux corresponding to the changes in the continuum. On the other hand, in the same project, they 
found that this correlation is not true for the sources if the continuum changes by more than $\sim$ 70\%.  
They suggest that for cases in which the $\mathrm{H\beta}$ flux does not correspond to the continuum, 
light travel time effects may have to be included. 
Furthermore, there is still no absolute conclusion on the correlation between the emission lines width and the line luminosity. 
An anti-correlation was found in a sample of 85 quasars by \cite{1994ApJ...423..131B}, 
with single-epoch spectroscopic data. On the other hand \cite{2005ApJ...633..638W} 
found that there is a correlation between the emission lines width and the line luminosity. 
They conclude this from a spectroscopic sample of 315 quasars.

\section{Paper outline and sample selection}

In this paper we investigate the continuum and narrow line variability of a small sample of objects over a 
good fraction of a decade.
The spectra cover a major fraction of the entire optical wavelength domain from the blue to the red,
including several prominent lines, and also have an appreciable signal to noise ratio in the continuum emission.
For such a study it is essential that the spectra are well calibrated such that they can be compared to each other. 
It is also important to correct for iron emission in order to extract proper values for the
line luminosities of species other than iron.
As the data has been taken with different instruments and different effective apertures on the sky it 
is furthermore essential to make sure that aperture effects, 
which may occur due to slit losses, or from extended and spatially resolved
line or continuum emission from the host galaxies, do not contaminate the results of the variability assessment.
In the literature one rarely finds appropriately calibrated or sufficiently described data sets that are
suitable for long term variability studies.
Hence we made the effort to collect and combine a representative data set.
This comprises a total of 18 sources (see summary in Sec. \ref{subsec:variab}).
For 8 objects we obtained and/or performed the detailed 
analysis over the available optical wavelength range using data from the LBT and other observatories
(see Sect.~\ref{sec:MODS} and Tab.~\ref{all_sources1}).
These 8 sources were selected on the basis of their observability (given the allocated observing dates),
their brightness (such that with the chosen integration times a decent signal to noise could be reached that allows
comparison to public survey data), and the fact that a previous first epoch spectrum had been taken typically 5-10 years ago.
Based on their [MgII] absorption, the 8 sources include 3 loBAL QSOs. However, given that their spectra also contain
prominent broad emission lines, we decided to include them in the characterization of the BLQSOs/Sy1 variability. 
In order to reach our goal of studying the difference in the variability properties of BLQSOs/Sy1 and NLSy1/S2 galaxies,
it was clear that literature data on additional sources had to be added to enlarge the statistical basis. 
Hence, all conclusions drawn in this paper are based on the resulting larger samples of sources.

For 5 sources (3 NLS1 and only 2 BLS1) we obtained spectroscopic information from the literature.
For a further 5 sources (2 NLS1 and 3 BLS1) we found sparser but suitable line and continuum information in the literature.
A detailed description of the observations and data reduction 
is given in Sect.~\ref{sec:obs_red}, followed by a description of the methods used for our 
analysis in Sect.~\ref{sec:analy}. In Sect.~\ref{sec:results} we present the results 
for all galaxies that we used in our study.

The sample also comprises sources with a range of identifications. 
For the purpose of this investigation we combined sources that 
show very prominent broad emission lines, i.e. Broad Line Seyfert~1 and 
Quasi Stellar Objects (in the following BLS1 and QSO)
in addition to the narrow lines on which our investigation focuses.
 We also put Seyfert~2 and Narrow Line Seyfert~1 (S2\footnote{One of our sample 
sources is a composite HII/S2 galaxy
- but see discussion in Sect.~\ref{sec:j0938}} and NLS1) 
in one group as their spectra are dominated by narrow line emission.
An additional justification for this combination is based on the fact that 
while the unified model of Seyfert galaxies suggests that there are hidden broad-line regions (HBLRs) 
in all Seyfert 2 galaxies, there is increasing evidence for the presence of a subclass of 
Seyfert 2 sources lacking these HBLRs \citep{ZhangWang2006, Haas2007}.
In these cases one cannot exclude that we have a free sight to the nucleus despite of
the fact that the source is classified as a S2 galaxy.
We also refer here to the discussion in the review on NLS1 galaxies by \cite{Komossa2008}.
Hence, if one classifies sources based on the presence of absence of pronounced broad line emission, 
then S2 and NLS1 may be more comparable to each other than e.g.  BLS1 with NLS1 sources.
In the discussion in Sect.~\ref{sec:discussion} and ~\ref{subsec:variab}
we analyze the observed degrees  of continuum and narrow line variability.
In retrospect we found that the above described combination of sources is indeed 
justified by their different variability characteristics.

Clearly, at some point such an investigation needs to be performed using larger samples.
Since an appropriate baseline for variability studies is several years, such an effort needs time and can 
only be provided in the near future.
However, to study first order effects, the number of sources needs to be only sufficiently large
to separate the median or mean properties within the statistical uncertainties.
As we outline in the paper, this can already be done with the current, representative sample presented here.
In fact, the interpretation we present in Sect.~\ref{subsec:accretion}
may also be taken as a prediction of the effect
that the degree of the continuum  variability is indeed reverberated by the degree of narrow line variability.
This prediction was motivated by the results of the analysis of the small but representative sample 
we present in this paper.
A conclusion is given in Sect.~\ref{sec:conclusion}.

\section{Observation and Data Reduction}
\label{sec:obs_red}
In this section, we describe the telescopes and instruments we used for the spectroscopic
observations and the procedures followed for the data reduction.
We also use complementary data from the literature and  public archives like the 
seventh data release of the SDSS \citep{2009ApJS..182..543A}.

\subsection{Multi-Object Double Spectrograph (MODS)}
\label{sec:MODS}
For our observations we used the MODS spectrograph at the Large Binocular Telescope (LBT)
\citep{2010SPIE.7735E...9P,2012SPIE.8446E..0GP}.
The LBT is located on Mount Graham in the Pinaleno Mountains southeastern Arizona, USA. 
The MODS spectrograph consists of a pair of identical double beam blue \& red optimized optical 
spectrographs (MODS1 \& MODS2) which work as individual spectrographs. The instrument can be used 
for imaging, long-slit, and multi-object spectroscopy. The charge-coupled device (CCD) detector of 
MODS is an e2vCCD231-68 $8k \times 3k$ with a 15 $\mu$m pixel pitch. 

We use MODS in long-slit mode with a slit width of 0\farcs 8. The spectrograph operates between 
3200-10500 \AA ~with a spectral resolution of R=$\lambda/\delta\lambda\approx2000$ 
(i.e., $\sim150\ \mathrm{km}\ \mathrm{s}^{-1}$) where the full spectral range of the 
grating spectroscopy is split into a blue and red channel, 
which are 3200-6000 \AA ~(blue) and 5000-10500 \AA ~(red). 
All objects which we targeted with our LBT observing program are listed in 
Tab.~\ref{all_sources1}.

\subsection{The Sloan Digital Sky Survey}
\label{sec:SDSS}
The Sloan Digital Sky Survey (SDSS) is a sensitive photometric and spectroscopic public survey which started science operation in May 2000 and covers $10^4$ square degrees of the celestial sphere in the northern sky \citep{2002AJ....123..485S}. 
The survey uses a dedicated 2.5-m wide-angle optical telescope mounted at Apache Point Observatory 
in New Mexico, United States 
(Latitude 32\degr  46\arcmin  49\farcs30 N, Longitude 105\degr 49\arcmin 13\farcs50 W, Elevation 2788m). 
This instrument provides images and photometric parameters in five bands ($u$, $g$, $r$, $i$, and $z$) 
with an average seeing of  1\farcs 5  and down to a limiting magnitude of $\sim 22.2$ in $r$ band. 
Spectroscopy of selected objects is done through 3\arcsec fibers with calibration uncertainties of the order of 2\%. 
The spectral resolution R of the instrument is about $R\sim 2000$, covering a wavelength range between 
3800 and 9200 \AA. 

In this research we used the SDSS active galaxies (Seyfert \& Quasar) Catalogs based on the
seventh data release of the SDSS \citep{2009ApJS..182..543A}. 

All our project targets (except J035409.48+024930.7 and J015328+260939 which are not covered by the SDSS survey)
can be found in different SDSS catalogs releases 
(DR5, DR6, DR7, DR8, and DR9) \citep{2000AJ....120.1579Y, 2002AJ....123..485S, 2009ApJ...692..758G, 2006AJ....131.1163S, 2008ApJS..175..297A, 2009yCat.2294....0A, 2009ApJS..182..543A,  2011yCat.2306....0A, 2012AJ....143..119I}. 

Typical magnitudes of SDSS sources e.g. in the g-band are in the range of 14.47 $\leqslant \mathrm{g} \leqslant$ 19.21, 
the FWHM of the combined bright narrow and broad lines typically range from 
about 300 to 3000 $\mathrm{km}\ \mathrm{s}^{-1}$. 

In this paper we analyze the optical spectra
of eight galaxies, one is classified as composite HII/S2, 3 sources are classified as S1 (BLS1 \& NLS1), and 4 sources are classified as QSOs. 
Further information about the analyzed objects is presented in Tab.~\ref{all_sources1}.
In Figs.~\ref{fig:subtraction} and ~\ref{fig:subtraction_rest}, we show images of the sources and a 
nearby reference star extracted from the same image frame. 
 
\subsection{Data reduction, wavelength calibration, and flux calibration}
\label{sec:data_reduc}
Here we describe the data reduction of our spectroscopic observations, wavelength calibration, 
flux calibration, and the complementary data from public archives that has been used in our analysis.

For reducing the data of the sources listed in Tab.~\ref{all_sources1}, we used the python software modsCCDRed, 
which is provided by the instrument team. This software package gives us the possibility to create bias correction, 
flat fielding, normalization, and other standard steps of data reduction.

First we create normalized spectral flat field frames for each channel (blue and red) 
through the following steps: 

\begin{itemize}
 \item bias correction of the flat fields images;
\item median combination of the bias-corrected flats;
\item interpolation of the bad columns using the bad pixel lists for the detector;
\item elimination of the (wavelength dependant) color term to produce a normalized ``pixel flat''.
\end{itemize}

More information on producing the normalized calibration frames can be obtained from the instrument manual  
\footnote{\url{http://www.astronomy.ohio-state.edu/MODS/Manuals/MODSCCDRed.pdf}}.
After generating the flat, all bias-corrected science data were flat fielded. 
The science frames were then wavelength-calibrated
using a wavelength-calibration map generated from lamp
(argon, neon, xenon and krypton)  data, as provided by the instrument.
In this step, we used the reduction package \textsc{Iraf} to identify the lamp lines and then transformed the 
science frames accordingly. Furthermore, we tested the wavelength-calibration 
by using an OH-sky line atlas \citep{1996PASP..108..277O,1997PASP..109..614O} and found 
that both calibration methods (skylines and lamp) are in good agreement with an uncertainty $\leqslant 3 \AA$.

\begin{figure}
\begin{center}
\includegraphics[width=\linewidth]{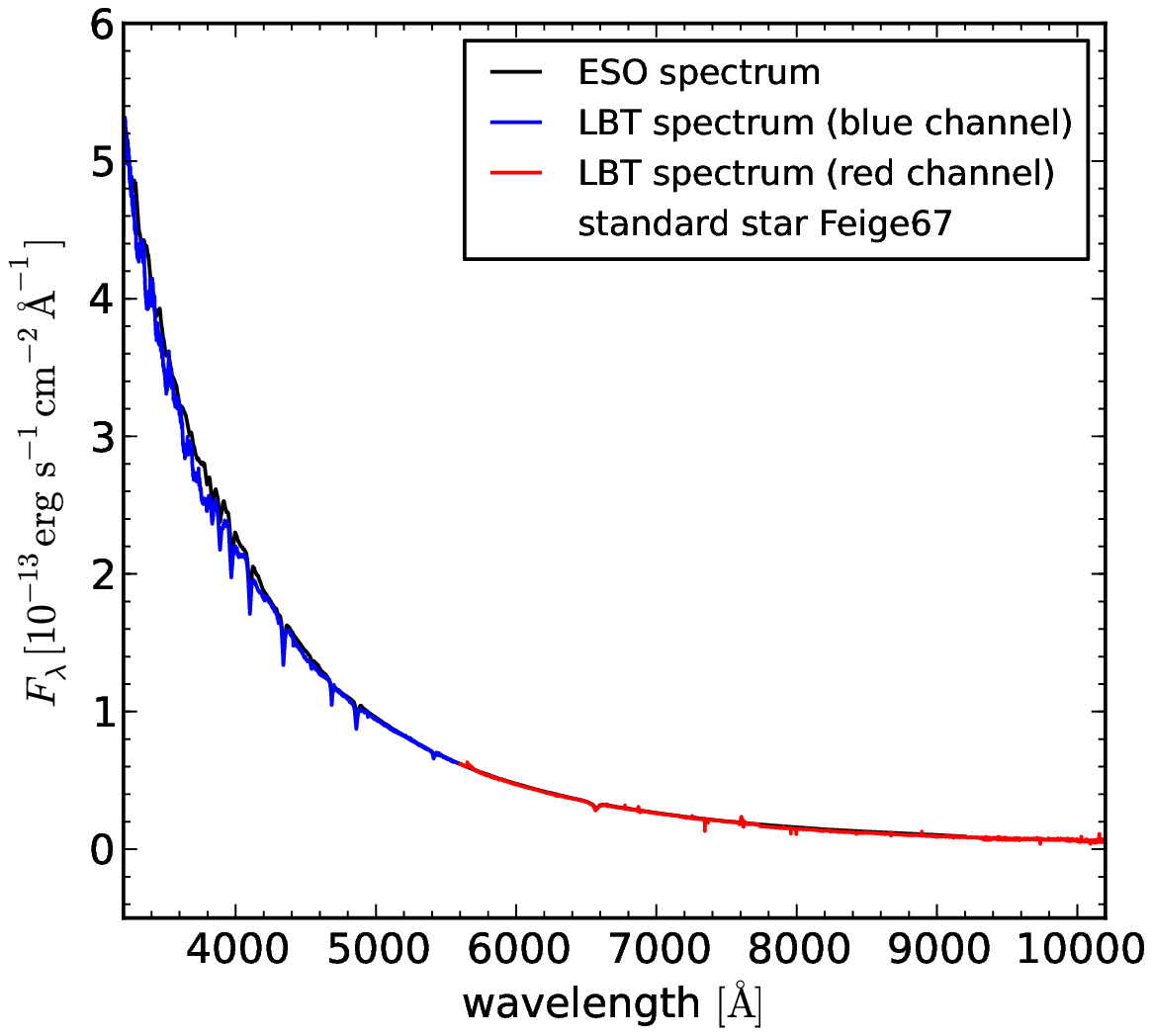}\\
\includegraphics[width=\linewidth]{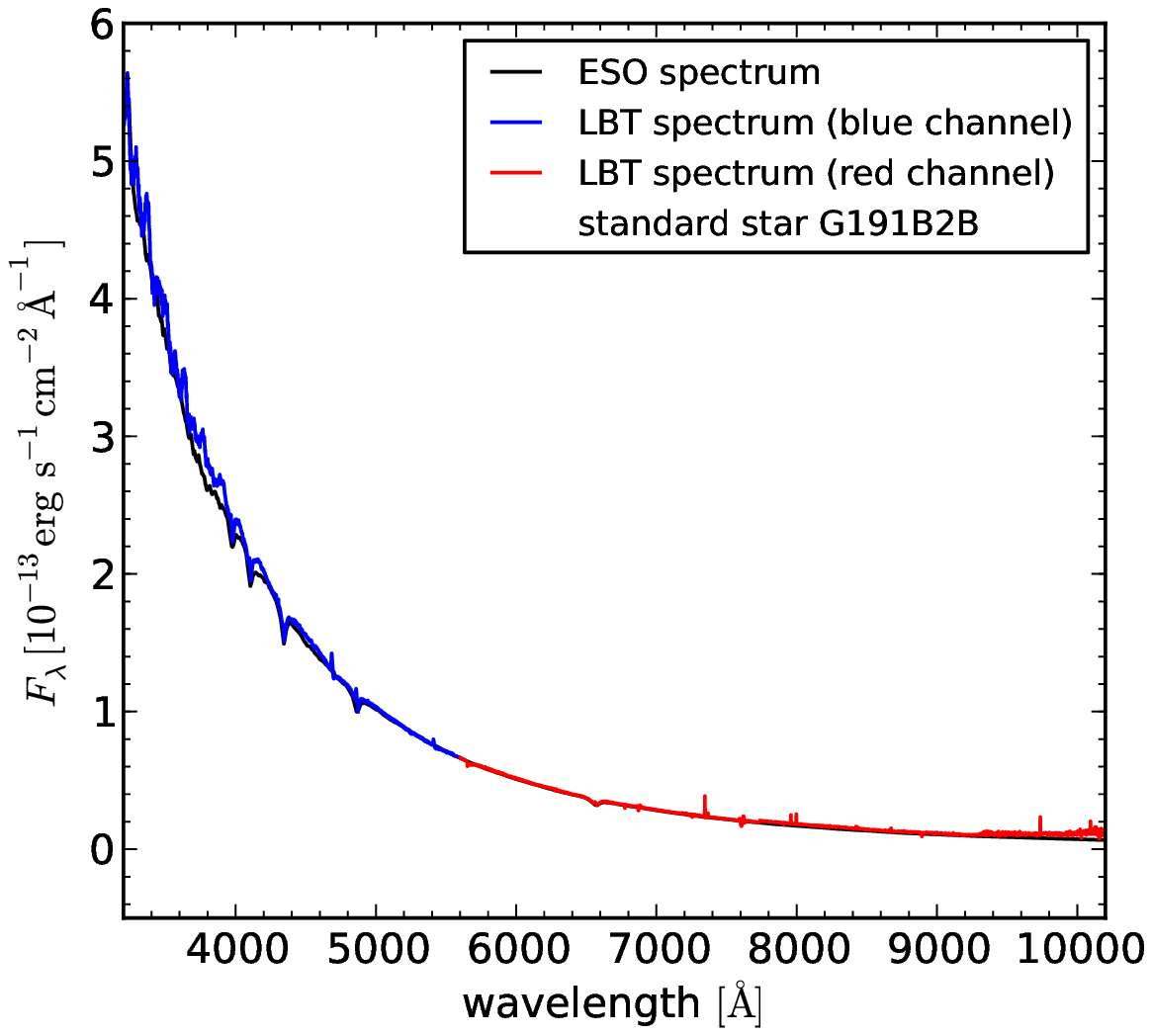}\\
\end{center}
\caption{
The standard stars Feige67 and G191B2B.
}
\label{standard_star}
\end{figure}

For the next step, we flux-calibrated the one-dimensional spectra making use of two 
standard stars: G191B2B and Feige67 for which data was taken under 1.3'' seeing conditions.
Both standard stars  result in a consistent and independent calibration.
For the final data product we used both stars for calibration.
We used reduction routines on two different software platforms: IDL and IRAF.
We extract the one-dimensional spectra from the two-dimensional 
science frames of the two calibration star using routines in the \textsc{Iraf} package.
For test purposes we calibrated the two stars with each other and compared the results 
with published data.
We also downloaded spectra of these two standard stars from the European Southern Observatory 
\footnote{\url{http://www.eso.org/sci/observing/tools/standards/spectra/}} 
and compared them with our results (see Fig.~\ref{standard_star}). 
Calibrating the science objects with the two reference stars independently resulted in spectra 
that agreed very well with each other within an average uncertainty of 2\%  and 3\% in the blue and red channel, respectively.
We performed that calibration for each source for the red and blue channel and combine 
these spectra to a single spectrum per source using \textsc{Iraf}. 
In this process, we found that there are no significant continuum offsets in the overlap area between the red and blue side.
While this confirms the consistency of our calibration, the uncertainties
in comparing data between different telescopes and instruments is more of the
order a few percent, \citep[see e.g.,][]{Peterson2002}
i.e.  of the same order as reached for the SDSS data releases.

In order to ensure a sufficient inter-calibration accuracy between telescopes and instruments
we applied a correction for slit losses.
We observed our target sources (including the two calibrated stars; see above)
in typically 1.2 to 1.4 arcsec seeing through a 0.8 arcsec slit. 
For the two extreme cases we applied a seeing correction based on Fig.~\ref{slitlosses}.
For J0153 taken in 1.90'' seeing we scaled the LBT fluxes up by a factor of 1.35.
For J1203 taken in 0.82'' seeing we scaled the LBT fluxes down by a factor of 0.70 SDSS.

In order to correct for a time variable slit loss contribution we proceeded in the following way:
For each target we have 2 to 5 exposures such that we have a good statistical estimate
on the flux loss due to the combination of variable seeing and misplacement of the slit.
In Fig.~\ref{slitlosses} we show a number histogram of the intensity drop measured in the
[OIII] $\lambda$5007, [OII] $\lambda$3727, H$\beta$ or H$\alpha$ line
in the corresponding red and blue channel
for the fainter exposures with respect to the brightest once.
To first order we correct for this effect by adjusting all exposures to the flux level of the
brightest exposure per source.
Under the assumption that for the brightest exposure the slit was always centered on the source
no additional correction for slit losses is necessary
(i.e. a factor of unity; case $\alpha$ in Fig.~\ref{slitlosses}).
However, if we assume that no significant systematical misplacements happened
and that the statistics for slit losses for the brightest exposures is similar to the statistics of the
fainter exposures then an unfavorable and unlikely case will be that all brightest exposures
were taken under conditions of a mean slit loss (case $\beta$ in Fig.~\ref{slitlosses}).
For a final second order correction for slit losses we assumed a case between case $\alpha$ and case $\beta$
and chose the mean between unity and the factor for mean losses of 1.17$\pm$0.10, 
i.e. a second order correction factor of 1.08, after having applied the first order correction.
Judging from Fig.~\ref{slitlosses} the uncertainty on this correction for the time variable slit losses 
is probably of the order of 5\% for the 1.2'' to 1.4'' seeing cases. 
For J0153 with 1.9'' this correction is probably overestimated by $<$5\% 
For J1203 with 0.82'' seeing our correction may be underestimated by 5-10\% especially if one takes into account that
for a better seeing slit positioning can also be done more reliably.
Given the seeing and aperture combinations for the Beijing and Hiltner telescope measurements 
(see section \ref{sec:j0153} and \ref{sec:j0354})
as well as the high SDSS calibration quality (see section~\ref{sec:SDSS})
no seeing/slit loss corrections were applied to these data.

As we flux calibrate our observed spectra with the flux calibration stars following the procedure presented above
and show that the galaxies are all
very compact (see section 3.2 and Fig.1 and A.2 and 
and section \ref{sec:j0938} for the marginally extended source J0938)
we assume the inter-calibration uncertainties to be less than 10\%.

 \subsection{Magnitude measurements}
\label{sec:mag_meas}
In the SDSS survey magnitudes are derived  using different methods
(Model, Cmodel, Petrosian, and PSF). For our work we used magnitudes derived via the PSF method 
in which a Gaussian model of the PSF is fitted to the object (dominated by a compact PSF like source). 
The method also accounts for the variation of the PSF across the field and was hence considered as 
most suitable for our study.

We compared the flux density value of the photometric observations\footnote{We use the 
Gemini flux density/magnitude converter (\url{http://www.gemini.edu/?q=node/11119})
to convert the magnitudes derived from the photometric images in the $u$,$g$,$r$,$i$, and $z$-bands into 
physical units (flux density $\mathrm{{F}_{\lambda}}$ in erg s$^{-1}$ cm$^{-2}$ \AA$^{-1}$).}
with those obtained from spectroscopic observation for each source. 
In this way we can get an estimate on the degree of variability that is present in the nuclei of these objects. 

\begin{figure*}
\includegraphics[width=0.3\linewidth]{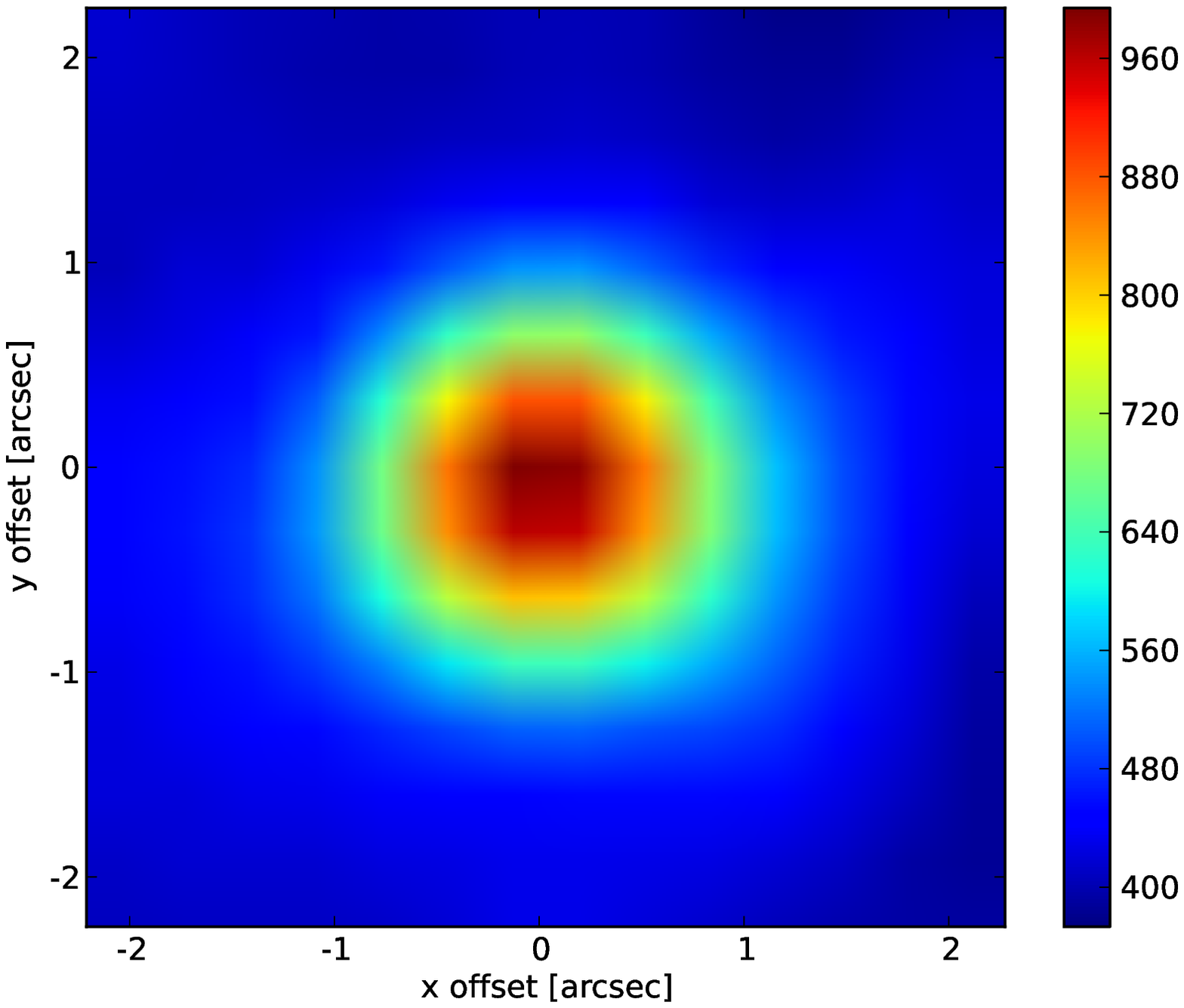}
\includegraphics[width=0.3\linewidth]{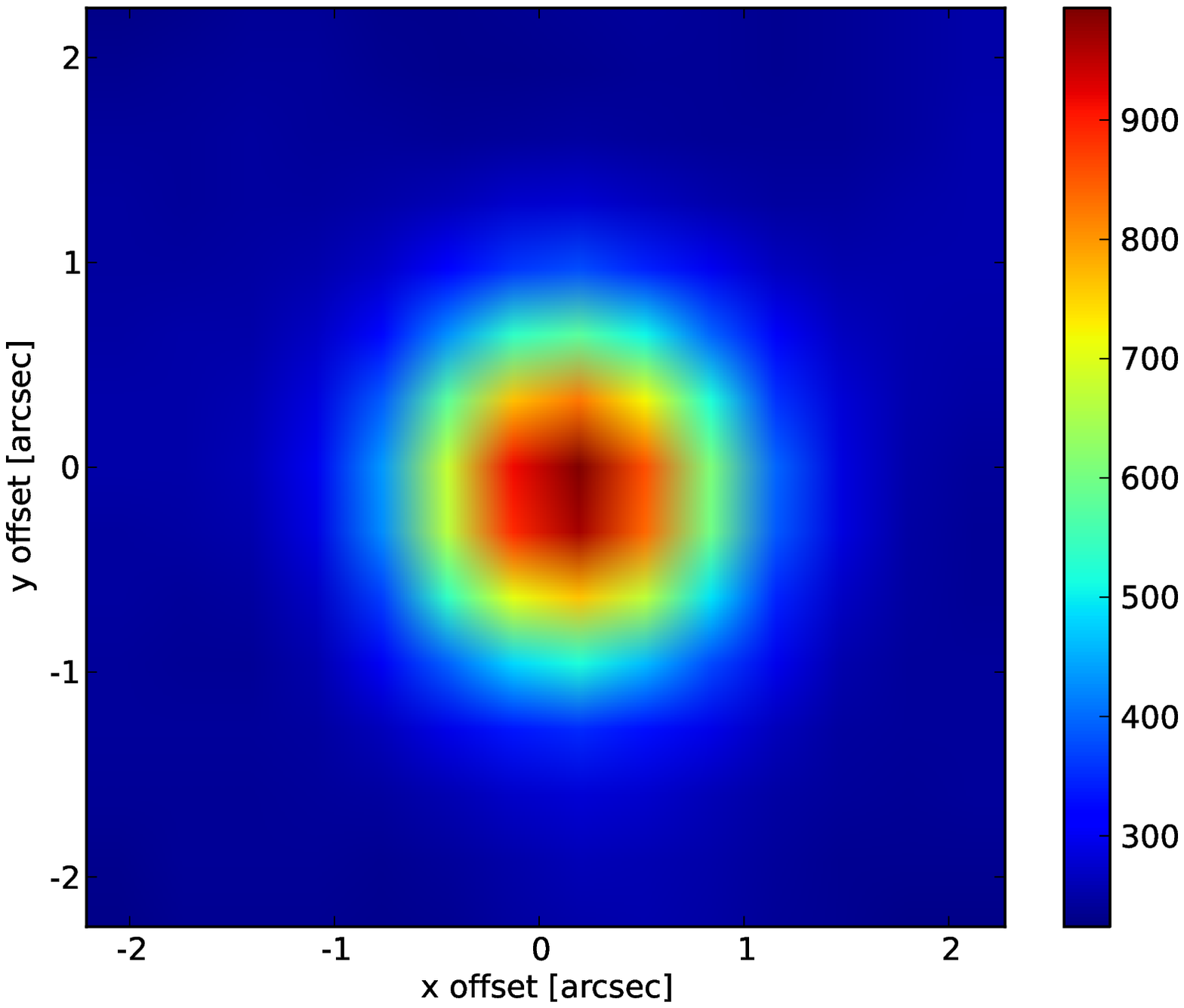}
\includegraphics[width=0.3\linewidth]{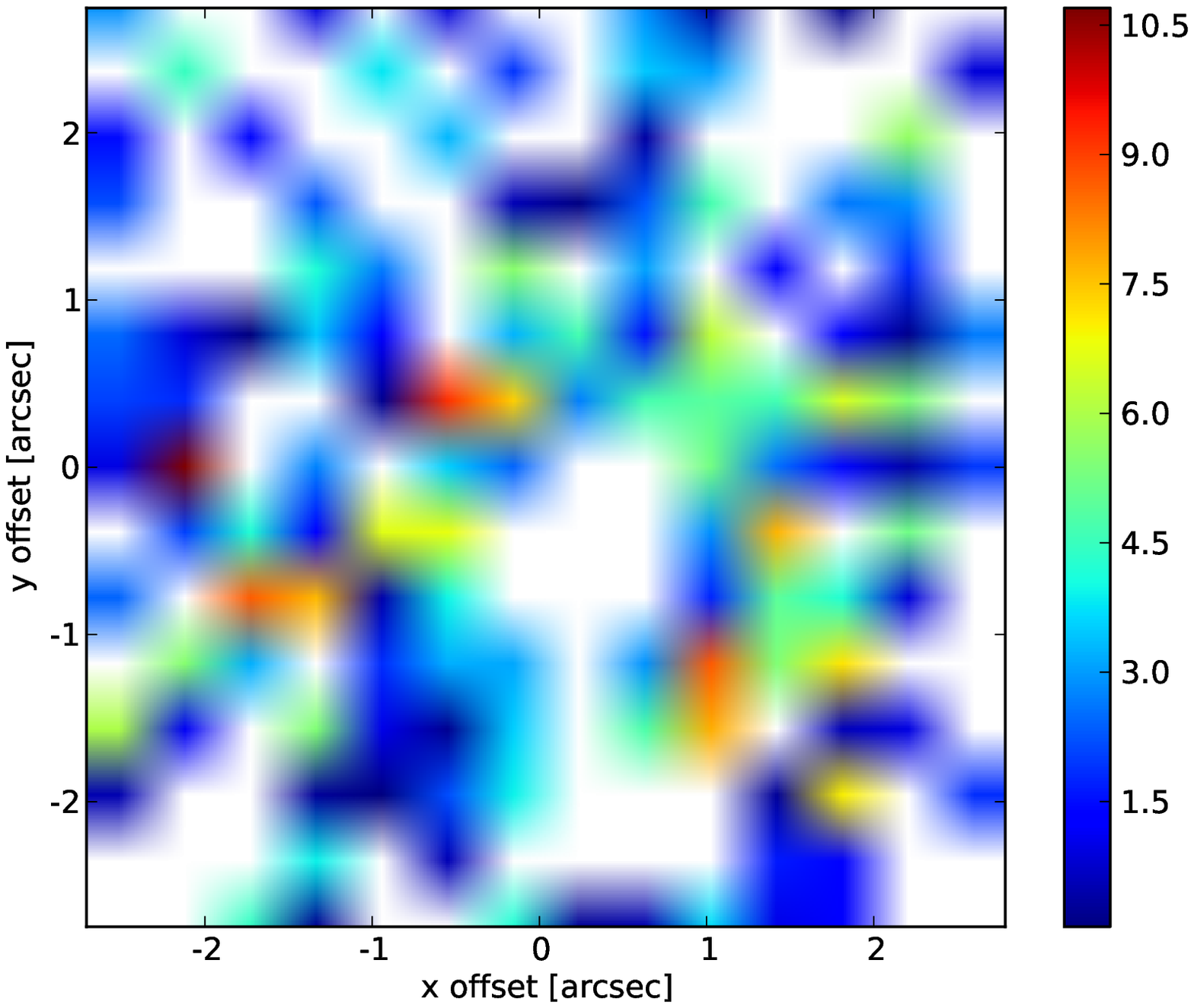}

\caption{The plot shows \emph{from left to right} an SDSS image of the galaxy J1203, 
a star/point source from the same SDSS frame (coordinates are listed in Tab.~\ref{all_sources2}) 
and the residuum that is left after subtracting the scaled star from the galaxy. 
The same plots for the other galaxies can be found in the appendix.}
\label{fig:subtraction}
\end{figure*}

In our LBT observations we used a long slit with aperture of 0\farcs 8, while for the SDSS observations a spectroscopic fiber 
with radius $3\arcsec$ was used. To demonstrate that aperture effects between slit and fiber measurements are negligible,
 we show SDSS $z$-band images of the galaxies together with images of nearby stars from the same frame 
(Figs.~\ref{fig:subtraction} and~\ref{fig:subtraction_rest}). 
We scale the stars to the flux level of the galaxy and subtract the frames from each other. 
In all eight cases, no emission is left. This shows that the sources are all dominated by emission 
from an unresolved combination of a stellar bulge and an AGN. 

\begin{landscape}
\begin{table}
\begin{center}
\caption{Coordinates, redshifts, classification and observing parameters for all sources.}   
\begin{tabular}{l c c c c c c c c c c}      
\hline
name &RA        &Dec.                         &Redshift (z) &Classi.&\multicolumn{2}{c}{Obs. date 1}&Obs. date 2&Photo.&$D_\mathrm{L}$&Ref.\\ 
&  &                    & &&&&&obs. date&\\ 
       &(hh mm ss)&($\degr$ $\arcmin$ $\arcsec$)&             &       &Telesc.&Da.(M Y)         & Da.(M Y)    &  Da.(M Y)   &Mpc           &           \\
\hline  
SDSS J120300.19+162443.8     &12 03 00.1&16 24 43.8&$0.16552\pm0.00001$& NLS1    &SDSS   &04 2007&02 2012&06 2005&794.5 &1 \\
SDSS J093801.63+135317.0     &09 38 01.6&13 53 17.0&$0.10056\pm0.00001$& HII/S2  &SDSS   &12 2006&02 2012&01 2006&463.2 &2 \\ 
SDSS J034740.18+010514.0     &03 47 40.1&01 05 14.0&$0.03149\pm0.00003$& NLS1    &OHP$^a$&09 2003&02 2012&11 2001&138.1 &3,4 \\
SDSS J115816.72+132624.1     &11 58 16.7&13 26 24.1&$0.43966\pm0.00035$& QSO     &SDSS   &05 2004&01 2012&03 2003&2429.5&5,6 \\
SDSS J080248.18+551328.9     &08 02 48.1&55 13 28.8&$0.66406\pm0.00035$& QSO     &SDSS   &01 2005&01 2012&11 2003 &3993.9&5,7\\
SDSS J091146.06+403501.0     &09 11 46.0&40 35 01.0&$0.44121\pm0.00099$& QSO     &SDSS   &01 2003&02 2012&12 2001&2439.7&5,6 \\
2MASX J03540948+0249307$^*$  &03 54 09.4&02 49 30.7&$0.03600\pm0.00008$& BLS1 &Hiltner$^b$&10 2004&02 2012&---   &158.4 &3,8 \\
GALEXASC J015328.23+260938.5$^*$&01 53 28.2&26 09 39.1&$0.32640\pm0.00127$& QSO &Beijing$^c$&02 2002&01 2012&--- &1711.9&9,10\\                                   
\hline                                                                        
\label{all_sources1}
\end{tabular}
 \begin{flushleft}
\it{\textbf{References for redshifts and positions:} (1)~\citet{2011yCat.2306....0A}; (2)~\citet{2011ApJ...735..125S}; (3)~\citet{2006A&A...455..773V}; (4)~\citet{1991ApJS...75..297H}; 
(5)~\citet{2010MNRAS.405.2302H}; (6)~\citet{2010ApJ...714..367Z}; (7)~\citet{2009ApJ...692..758G}; (8)~\citet{1981ApJ...243L...5C}; (9)~\citet{2002AJ....124..100D}; 
(10)~\citet{2002ApJ...581...96Z}.\\
\textbf{Notes:} The 3 QSOs (J1158, J0911, J0802) contained in the SDSS catalogue have been classified as low-ionization broad absorption-line 
(loBAL) due to a prominent [\ion{Mg}{ii}] absorption in their spectra. The second spectroscopy observation have always taken by LBT. 
The photometric observations were always taken by SDSS. For the final two sources in the table no SDSS photometric observations
are available.\\
$^*$ Not covered by SDSS survey\\
$^a$ Telescope of Observatoire de Haute-Provence (OHP)\\
$^b$ Hiltner Telescope\\
$^c$ Beijing Observatory\\
}
      \end{flushleft}
\end{center}
\end{table}
\end{landscape}

\begin{table*}
\begin{center}
\caption{Comparison of seeing conditions for the sources for which we analyzed the spectra.}   
\begin{tabular}{l c c c c c c c c c c}      
\hline
Sources     &LBT seeing& \multicolumn{2}{c}{Comparing Spectro.}&SDSS photometric\\
            &          & Inst.& seeing                      &   Median Seeing (r-band)\\
            & \arcsec  &       & \arcsec                      & \arcsec  \\
\hline  
J1203      &0.82&SDSS&  1.48    &1.4\\
J0938      &1.30&SDSS&  1.69    &1.4\\
J0347      &1.40&OHP& 2.5       &1.4\\
J1158      &1.40&SDSS&  1.22    &1.4\\
J0802      &1.20&SDSS&  1.49    &1.4\\
J0911      &1.20&SDSS&  1.23    &1.4\\
J0354      &1.20&Hiltner& 2.5   &--\\
J0153      &1.90&Beijing&2.5    &--\\                        
\hline                                                                        
\label{seeing}
\end{tabular}
\end{center}
\end{table*}

\begin{table*}
\begin{center}
\caption{The observed continuum flux density variability between LBT and SDSS/OHP of J0938, J1203, J1158, J0911,
J0802, and J0347 from the photometry and spectroscopy aspect. $^{(a)}$For J0347 first epoch spectroscopy obtained by OHP.}
\begin{tabular}{l c c c c c c}
\hline
 Sources   &Filt.   &Wa. &photometry &Emission line  &spectroscopy & spectroscopy \\
  &  &  &  SDSS&  contribution in &SDSS$^{(a)}$ &   LBT \\
  &  &     \AA  &[$10^{-16}\,\mathrm{erg}$ &spectral filters&[$10^{-16}\,\mathrm{erg}$&[$10^{-16}\,\mathrm{erg}$\\
  &  &     &$\mathrm{s}^{-1}\ \mathrm{cm}^{-2} \AA^{-1}$]&in \% of continuum&$\mathrm{s}^{-1}\ \mathrm{cm}^{-2} \AA^{-1}$]&$\mathrm{s}^{-1}\ \mathrm{cm}^{-2} \AA^{-1}$]\\
    \hline
 \multirow{5}{*}{J1203} & (u)  & 3543& $ 1.26\pm0.03 $& &  ---       &$2.04 \pm0.99$\\
     & (g)  & 4770   &$ 1.06\pm0.02 $&   9&$ 1.20\pm 0.19$ &             $1.54 \pm0.38$\\
     & (r)  & 6231   &$ 1.53\pm0.01 $&  25&$ 1.02\pm 0.31$ &             $0.87 \pm0.13$\\
     & (i)  & 7550   &$ 1.09\pm0.01 $&  33&$ 0.81\pm 0.21$ &             $0.61 \pm0.14$\\
     & (z)  & 9134   &$ 0.53\pm0.03 $&    &$ 0.65\pm 0.32$ &             $0.49 \pm0.13$\\
     \hline
  \multirow{5}{*}{J0938}    & (u) & 3543   &   $1.60 \pm0.03 $& & ---&$1.87 \pm0.31$\\
     & (g) & 4600   &$2.39 \pm0.02 $ &   2&$ 3.41\pm 0.19$ &             $2.66 \pm0.27$\\
     &(r)  & 6231   &$2.01 \pm0.02 $ &    &$ 3.30\pm 0.15$ &             $2.48 \pm0.31$\\
     &(i)  & 7625   &$1.67 \pm0.02 $ &   7&$ 3.05\pm 0.17$ &             $2.36 \pm0.41$\\
     &(z)  & 9134   &$1.50 \pm0.02 $ &    &$ 2.79\pm 0.27$ &             $2.12 \pm0.45$\\
     \hline
  \multirow{5}{*}{J0347$^{(a)}$} &(u)& 3543&$68.81\pm0.01$&& ---     &$52.79 \pm0.09$\\
    &(g)  & 4770    &$53.90\pm0.02$  &    6&$ 44.60\pm0.21 $ &             $60.38 \pm0.02$\\
    &(r)  & 6231    &$44.92\pm0.01$  &   36&$ 31.12\pm 0.15$ &             $39.23 \pm0.01$\\
    &(i)  & 7625    &$40.91\pm0.02$  &     &$ 27.71\pm 0.11$ &             $33.81 \pm0.02$\\
    &(z)  & 9134    &$26.93\pm0.01$  &     & ---             &             $26.39 \pm0.04$\\
\hline
 \multirow{5}{*}{J1158}  &(u)  & 3543   &   $5.98 \pm0.04 $ && ---   &$5.12 \pm0.63$\\
     &(g)  & 4770   &$4.29 \pm0.02 $ &  5&$ 4.62\pm 0.23$ &             $3.91 \pm0.33$\\
     &(r)  & 6400   &$2.87 \pm0.02 $ &   &$ 2.81\pm 0.14$ &             $3.31 \pm0.16$\\
     &(i)  & 7700   &$2.21 \pm0.02 $ &  6&$ 2.15\pm 0.16$ &             $2.65 \pm0.03$\\
     &(z)  & 9134   &$1.77 \pm0.03 $ &   &$ 1.48\pm 0.48$ &             $1.87 \pm0.07$\\
         \hline
 \multirow{5}{*}{J0802}  & (u)  & 3543&$0.41\pm0.06$&& ---           &$0.88 \pm0.49$\\
    & (g)  & 4770   &$0.97\pm0.01$   &   &$ 1.02\pm 0.12$  &             $1.37 \pm0.11$\\
    & (r)  & 6231   &$1.17\pm0.01$   &  2&$ 1.14\pm 0.11$  &             $1.49 \pm0.06$\\
    & (i)  & 7625   &$1.24\pm0.02$   &  3&$ 1.16\pm 0.20$  &             $1.71 \pm0.10$\\
    & (z)  & 9134   &$1.04\pm0.02$   &   &$ 0.97\pm 0.25$  &             $1.23 \pm0.29$\\
    \hline
 \multirow{5}{*}{J0911} &(u)  & 3543   &   $0.89 \pm0.04 $& &  ---   &$0.61 \pm0.27$\\
    &(g)  & 4770    &$0.86 \pm0.02 $ &  5&$ 0.67\pm 0.04$ &             $0.92 \pm0.22$\\
    &(r)  & 6200    &$0.85 \pm0.01 $ &   &$ 0.90\pm 0.04$ &             $1.10 \pm0.15$\\
    &(i)  & 7700    &$0.68 \pm0.02 $ &  7&$ 0.72\pm 0.03$ &             $0.96 \pm0.07$\\
    &(z)  & 9134    &$0.54 \pm0.03 $ &   &$ 0.56\pm 0.10$ &             $0.76 \pm0.19$\\
     \hline
\label{fig:variab}
\end{tabular}
\end{center}
\end{table*}

\begin{table*}
\begin{center}
\caption{Comparison of the apparent sizes of the sources and reference stars in the field.}
\begin{tabular}{c c c c c c c c c}
\hline
source &\multicolumn{2}{c}{Star} &\multicolumn{3}{c}{Star}&\multicolumn{3}{c}{Galaxy}\\
       &RA        &Dec.          & FWHM1 &FWHM2&Angle&FWHM1  &FWHM2&Angle\\
       &(hh mm ss)&($\degr$ $\arcmin$ $\arcsec$)&\arcsec&\arcsec &\arcsec&\arcsec&\arcsec       &\arcsec \\
\hline
J1203      &12 02 59.4&16 24 20.5&$0.95\pm0.03$&$1.03\pm0.03$&$1.24\pm0.56$&$1.26\pm0.15$&$1.09\pm0.09$&$10.97\pm0.29$\\
J0938      &09 38 22.3&13 49 54.4&$1.15\pm0.01$&$1.04\pm0.01$&$0.97\pm0.13$&$1.31\pm0.07$&$1.11\pm0.08$&$1.88\pm0.48$\\ 
J0347      &03 47 35.8&01 04 08.6&$0.99\pm0.01$&$1.31\pm0.01$&$1.81\pm0.15$&$1.09\pm0.01$&$1.35\pm0.01$&$0.87\pm0.16$\\
J1158      &11 58 14.2&13 25 57.0&$1.43\pm0.02$&$3.76\pm1.48$&$0.92\pm0.30$&$1.44\pm0.04$&$1.49\pm0.04$&$0.64\pm0.48$\\
J0802      &08 01 38.2&55 11 44.2&$0.57\pm0.01$&$0.60\pm0.01$&$0.78\pm0.17$&$0.67\pm0.04$&$0.52\pm0.03$&$49.44\pm0.46$\\
J0911      &09 11 47.7&40 34 38.9&$0.75\pm0.04$&$0.81\pm0.04$&$3.22\pm0.82$&$0.87\pm0.06$&$0.91\pm0.06$&$0.79\pm0.90$\\
\hline
\label{all_sources2}
\end{tabular}
\end{center}
  \begin{flushleft}
 \it{\textbf{Notes:} For each galaxy, we give the coordinates of the star that we subtracted from the galaxy image as 
 well as results of a Gaussian fit to galaxy and star. We used an elliptically shaped Gaussian function.}
  \end{flushleft}
\end{table*}

Furthermore, we fitted Gaussian functions to the images of the galaxies and the stars. 
The results are listed in Tab.~\ref{all_sources2}, supporting the finding that the galaxy emission 
is dominated by the unresolved point source. In Tab.~\ref{seeing} we list the seeing values for the different exposures.

\section{Analysis}
\label{sec:analy}

We developed a \textsc{Python} routine to manually fit the stellar continuum, the power-law 
contribution from the AGN, as well as the \ion{Fe}{ii} emission and subtract all of them. 
In the residual spectrum we can then fit the non-\ion{Fe}{ii} emission lines. In following we explain each step:

\subsection{Stellar continuum subtraction}
\label{sec:ste_cont}
To obtain an accurate measurement of the nuclear emission line
fluxes and the equivalent widths,  we have to subtract the stellar component of the host or its bulge component.
To remove the stellar component we fitted  a stellar population synthesis
model to the entire spectrum. The templates used are given by a sample spectrum
built by a population synthesis routine in \cite{2003MNRAS.344.1000B}.
The template for the young stellar population spectrum is taken 290 Myr after a starburst with
steady star formation rate over 0.1 Gyr and solar metalicity. For the intermediate-age population
we used template of a 1.4 Gyr old simple stellar population with solar metalicity.

\begin{table*}
\begin{center}
\caption{Continuum fits.}
\begin{tabular}{c c c c c c}
\hline
name    & instrument    & $A_V$ (stars)    & stars & powerlaw    & \ion{Fe}{ii}    \\
\hline
J1203     & LBT        & 0        & 60\%   & 40\% & 0\%        \\
          & SDSS       & 0        & 60\%   & 40\% & 0\%        \\
J0938     & LBT        & 1.5      & 90\%   & 10\% & 0\%        \\
          & SDSS       & 1.5      & 80\%   & 20\% & 0\%        \\
J0347     & LBT        & 1.2      & 57\%   & 20\% & 23\%        \\
          & OHP        & 0.8      & 72\%   & 13\% & 15\%        \\
J1158     & LBT        & 0        & 0\%    & 85\% & 15\%        \\
          & SDSS       & 0        & 0\%    & 85\% & 15\%        \\
J0802     & LBT        & 0        & 50\%   & 30\% & 20\%        \\
          & SDSS       & 0        & 52\%   & 33\% & 15\%        \\
J0911     & LBT        & 0        & 50\%   & 40\% & 10\%        \\
          & SDSS       & 0        & 45\%   & 45\% & 10\%        \\
J0354     & LBT        & 1.8      & 60\%   & 30\% & 10\%        \\
J0153     & LBT        & 0        & 62\%   & 23\% & 15\%        \\
\hline
\label{tab:contfit}
\end{tabular}
\end{center}
    \begin{flushleft}
 \it{\textbf{Notes:} For each observation we give the flux contributions of the stellar component, the AGN/powerlaw component, 
 and the \ion{Fe}{ii} template, as well as the extinction $A_V$ of the stellar component in the wavelength interval $5100\,\mathrm{\AA}$ to $5600\,\mathrm{\AA}$ (restframe wavelength).}
       \end{flushleft}
\end{table*}

In Tab.~\ref{tab:contfit} we show the flux contributions of the stellar component, the AGN/powerlaw 
component, and the \ion{Fe}{ii} template, as well as the extinction $A_V$ of the stellar component.
We measured the fraction of each component compared to the total flux in the (restframe) wavelength 
interval $5100\,\mathrm{\AA}$ to $5600\,\mathrm{\AA}$. 
The main purpose of the subtraction was to remove the \ion{Fe}{ii} emission lines from 
the spectrum, particularly those blended with other emission lines. 
Since in most galaxies there are no obvious stellar features, it is difficult to distinguish between 
different stellar populations. Therefore, we sum up the contributions of the intermediate-age 
and the young stellar population. Also the fit of the extinction $A_V$ is not reliable, but was 
only chosen to fit the continuum slope as accurate as possible. 
The data for the line variability as discussed in Sect.~\ref{sec:discussion}
includes all corrections derived from the fitting described here.

\subsection{\ion{Fe}{ii} subtraction}
\label{sec:fe}
\begin{figure}
\includegraphics[width=\columnwidth]{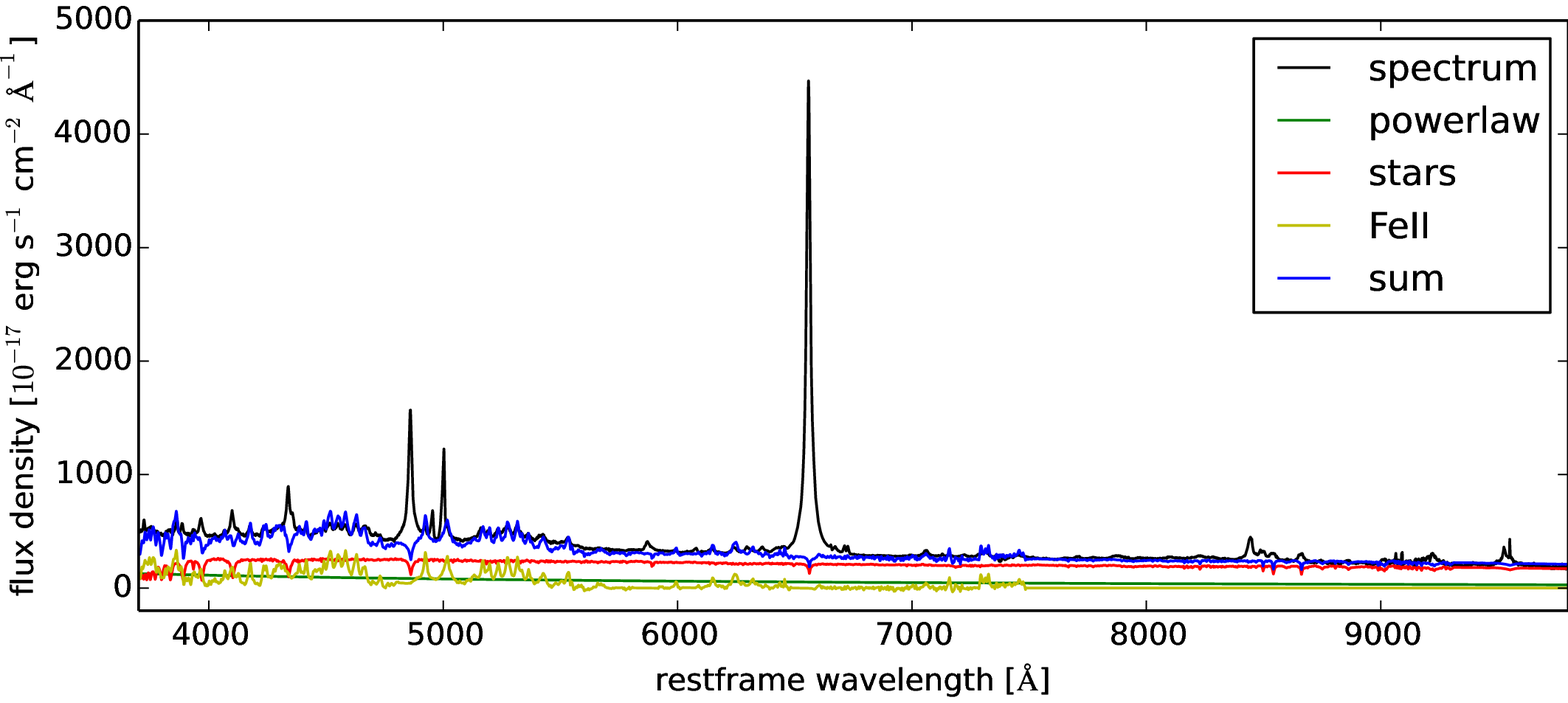}
\includegraphics[width=\columnwidth]{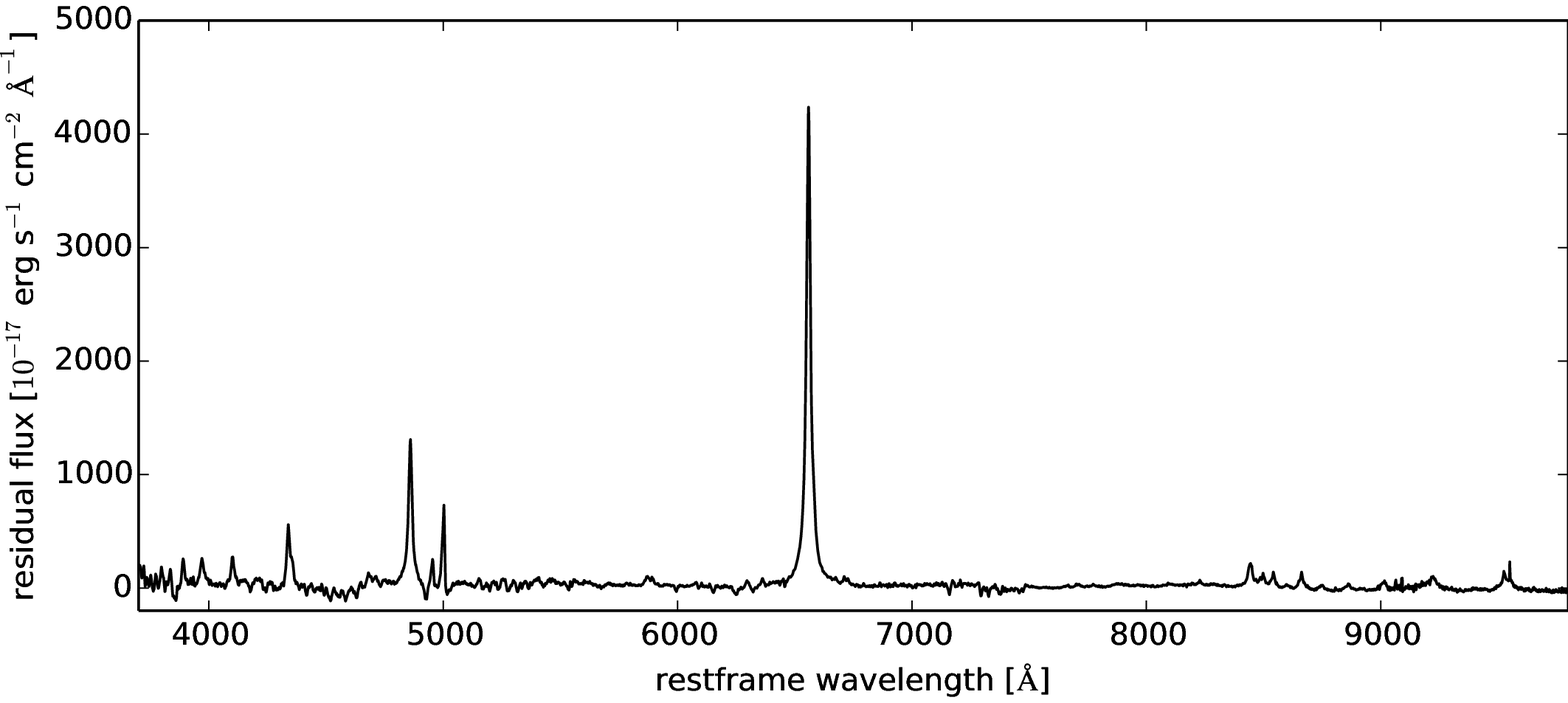}
\caption{
Results from the [\ion{Fe}{ii}] emission-subtraction. In the first plot, we show the spectrum of J034740.18+010514.0 together with a fit of the [\ion{Fe}{ii}] emission (yellow). In the second plot, shows the spectrum after subtraction of the [\ion{Fe}{ii}] emission.
}
 \label{fitting_feii}
\end{figure}

The optical emission-lines of \ion{Fe}{ii} are typical features of Seyfert~1 galaxies and
quasars. There is huge variety in the strength of these lines, from very clearly shaped and detectable
lines in some AGN spectra, all the way to very weak and imperceptible lines in other sources.
\cite{1992ApJS...80..109B} (in the following BG92) reported a strong inverse correlation between the
strength of \ion{Fe}{ii} emission-lines and the width of the broad $\mathrm{H\beta}$ line as well as the
strength of the [\ion{O}{III}] line. This result was confirmed by other groups
\citep[e.g.,][]{1997ApJS..113..245C, 2006ApJS..166..128Z, 2010ApJS..189...15K, 2011ApJ...736...86D}.
For our work we used the \ion{Fe}{ii} template from BG92, fitted and scaled
to our galaxy spectra as shown in Fig.~\ref{fitting_feii}.
Tab.~\ref{tab:contfit} shows that 
the contribution of \ion{Fe}{ii} in the given wavelength range is up to 23\% and 
a proper subtraction of \ion{Fe}{ii} is required, so
the fitted spectra were then subtracted from our spectra.
Finally, we obtained a spectrum corrected for both the host continuum (as mentioned in 
Sect.~\ref{sec:ste_cont}) and  \ion{Fe}{ii}.

\subsection{Emission-line fitting}
\label{sec:line_fit}
\begin{figure}
\includegraphics[width=\columnwidth]{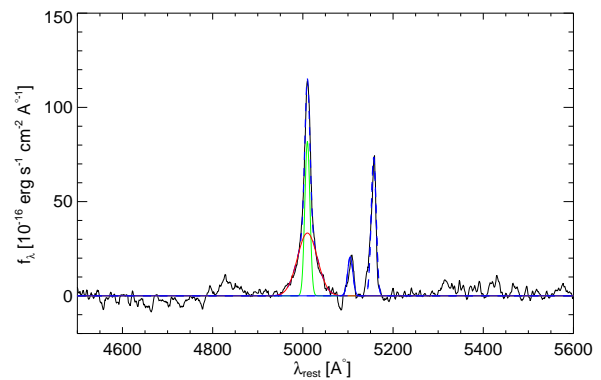}
\caption{The plot shows the fit of the H$\beta$ and $[OIII]$ emission line complex with multiple Gauss functions for J0347.}
\label{fitting_hb}
\end{figure}

To accurately measure the fluxes of emission lines, we used \textsc{mpfitexpr} \citep{2009ASPC..411..251M}. We fitted several Gaussians or a 
single Gaussian to the line profile, to fit narrow or broad components depending on the details 
of their shape or the presence of neighboring features. 
For example, to fit the H$\alpha$+\ion{[N}{ii]} complex, we used four Gaussians components, where the flux 
ratio of the \ion{[N}{ii]}$\lambda$ 6583 \AA, 6548\AA\ ~doublet is fixed to the theoretical 
value of 2.96. In addition, their widths are supposed to be identical. Likewise, for the complexes of 
H$\beta$+\ion{[O}{iii]} $\lambda$ 5007 \AA, we fitted three Gaussians components as shown in Fig.~\ref{fitting_hb}, 
and two Gaussian components if the H$\beta$ line is narrow. 
Additionally, for some broad lines we fitted a Lorentzian profile if the width of the lines
cannot be represented by Gaussian line profiles. 
The results of the line fitting are listed in Tab.~\ref{fluxes_fwhm}.

\section{Results}
\label{sec:results}
Redshifts and luminosity distances $D_\mathrm{L}$ \citep{1999astro.ph..5116H} are listed in Tab.~\ref{all_sources1}.
The luminosity distances $D_\mathrm{L}$ were calculated using the redshift of the source combined with the cosmology constants 
assuming a Hubble constant 
$H_0=70\ \mathrm{km} \ \mathrm{s}^{-1} \ \mathrm{Mpc}^{-1}$, and a standard cosmology with parameters
$\Omega_{\mathrm m}=0.3$, and $\Omega_\Lambda=0.7$ \citep[][which we use throughout the paper]{2003ApJS..148..175S}.\\
The observed continuum flux density variability of the sources derived from  LBT and SDSS data is summarized in
Tab.~\ref{fig:variab}. 

The main body of comparative variability studies of extragalactic nuclei is carried out without 
a detailed differentiation of contributions of different spectral components to the overall emission.
For the brighter QSOs this can be justified since the host contribution is small.
Such a spectral analysis requires a very high signal to noise in order to determine the exact nature of the 
stellar host contribution which in return is dependant on the modeling details of the dominant contributing 
stellar populations.
Given the inter-calibration quality for the newly presented 8 sources 
such a analysis shows that around 5500~\AA ~(restframe wavelength)
the stellar contribution to the integrated flux for the S2/NLS1 source is $\ge$50\% and for the QSOs it is
$\le$50\%. However, the quality of the data does not allow to study the variability of the power spectrum component alone,
since the details of the stellar population analysis required a much higher data quality.
Hence, the difference in the continuum variability properties between S2/NLS1 and QSO/BLS1 
presented in this paper is influenced by the presence of a stronger stellar contribution for the S2/NLS1. 
Details are given in column 5 of Tab.~\ref{fig:variab}.
If no value is given in column 5 then the contamination is below 5\%.

The variability of a larger sample (see section~\ref{subsec:variab})
is discussed with continuum and line variability extracted from spectra.
In order to avoid contamination from the host and the Fe-lines, the line variability 
is measured from spectra corrected for these contributions (see subsections above).
However, in order to be able to enlarge the sample with data from the literature, the continuum variability measurements 
do contain the stellar flux contribution.
The effects of this are discussed in section \ref{subsec:NLRresponse}.

The line fitting results are listed in Tab.~\ref{fluxes_fwhm}.
We plot all our spectra in the rest wavelength as shown in Fig.~ (\ref{spec_j0938}, \ref{spec_j1203}, \ref{spec_j0347}, etc.). 
We did that by applying $\lambda_{rest}= \lambda_{obs}/{1+\mathrm{z}}$.
Another correction was applied to the flux density (cosmological dilution), changing this value to the rest system via

\begin{equation}
F_\lambda\ rest = F_\lambda\ obs. \times (1+\mathrm{z})^{3} ~~~.
\end{equation}
All continuum flux densities and luminosities listed in the tables are derived as observed quantities (i.e. without corrections).
The listed line fluxes and luminosities have been corrected for FeII and stellar continuum as described above.

\subsection{SDSS J093801.63+135317.0}
\label{sec:j0938}
SDSS J093801.63+135317.0 (hereafter J0938) is discussed in the discovery of a population 
of normal field galaxies that have luminous and strong FHIL (forbidden high by ionized lines) 
and HeII$\lambda$4686 emission \citep{2011Ap.....54..340S}. The authors used the 
'Bolshoi Teleskop Alt-azimutalnyi BTA-6'  6-m diameter telescope in Russia and obtained 
spectra with 0.86\AA/px resolution. \cite{2011Ap.....54..340S} report, in their Table 2, 
the value of FWHM and flux for emission lines in the optical spectrum of J0938 
taken with two instruments with the same spectral resolution of $R \sim 2000$ at two different epochs. 
The first is from the SDSS telescope on 18.12.2006 and the second 
is from the BTA-6 on 16.04.2010. 

\cite{2013ApJ...774...46Y} present results of seven rare extreme coronal line emitting galaxies, 
and J0938 is also among these sources. These objects are reported by \cite{2012ApJ...749..115W}, and four of 
these galaxies (J0938 and three others) have a large variability in coronal line flux, making them good candidates
for tidal disruption events (TDEs). 
They detected a broad  He II$\lambda$4686 emission line in the spectrum of J0938. \cite{2013ApJ...774...46Y} 
find broad coronal and high-ionization lines that are superimposed on narrow low-ionization lines. 
They interpret this finding as indication that J0938 is a composite of a S2 nucleus and a star formation region.
In the SIMBAD catalogue the object is listed as an \ion{H}{ii} galaxy.
However, according to the [\ion{O}{i}]$\lambda$6300/H$\alpha$ versus [\ion{O}{iii}]$\lambda$5007/H$\beta$
diagnostic diagram in \cite{2012ApJ...749..115W} and \cite{2013ApJ...774...46Y} it's appearance to be that
the source J0938 is located in the region between \ion{H}{ii}
and S2 galaxies (here we adopt an HII/S2 composite nucleus classification) and belongs to a sample of sources
with spectra that are dominated by the interaction of a
super massive black hole with the nuclear environment
\citep{2011Ap.....54..340S, 2012ApJ...749..115W, 2013ApJ...774...46Y}.

The variability data for this source are consistent with the uncertain classification of the source as HII/S2
and support the presence of a variable nucleus.
In Tab.~\ref{fig:variab} we list the flux density measurements 
of J0938 for three epochs at different wavelengths
as obtained through our LBT observations
and via the spectroscopy and photometry results listed by SDSS.
We note that flux values of the continuum of LBT spectroscopy
are in good agreement with photometric data of SDSS at the
same wavelengths, while both are different to what can be derived
from SDSS spectroscopy. 
We notice that their are of order 30\% variations in line and continuum 
flux between the different epochs we use for our investigation.

\begin{figure*}
\begin{center}
\includegraphics[width=2.0\columnwidth]{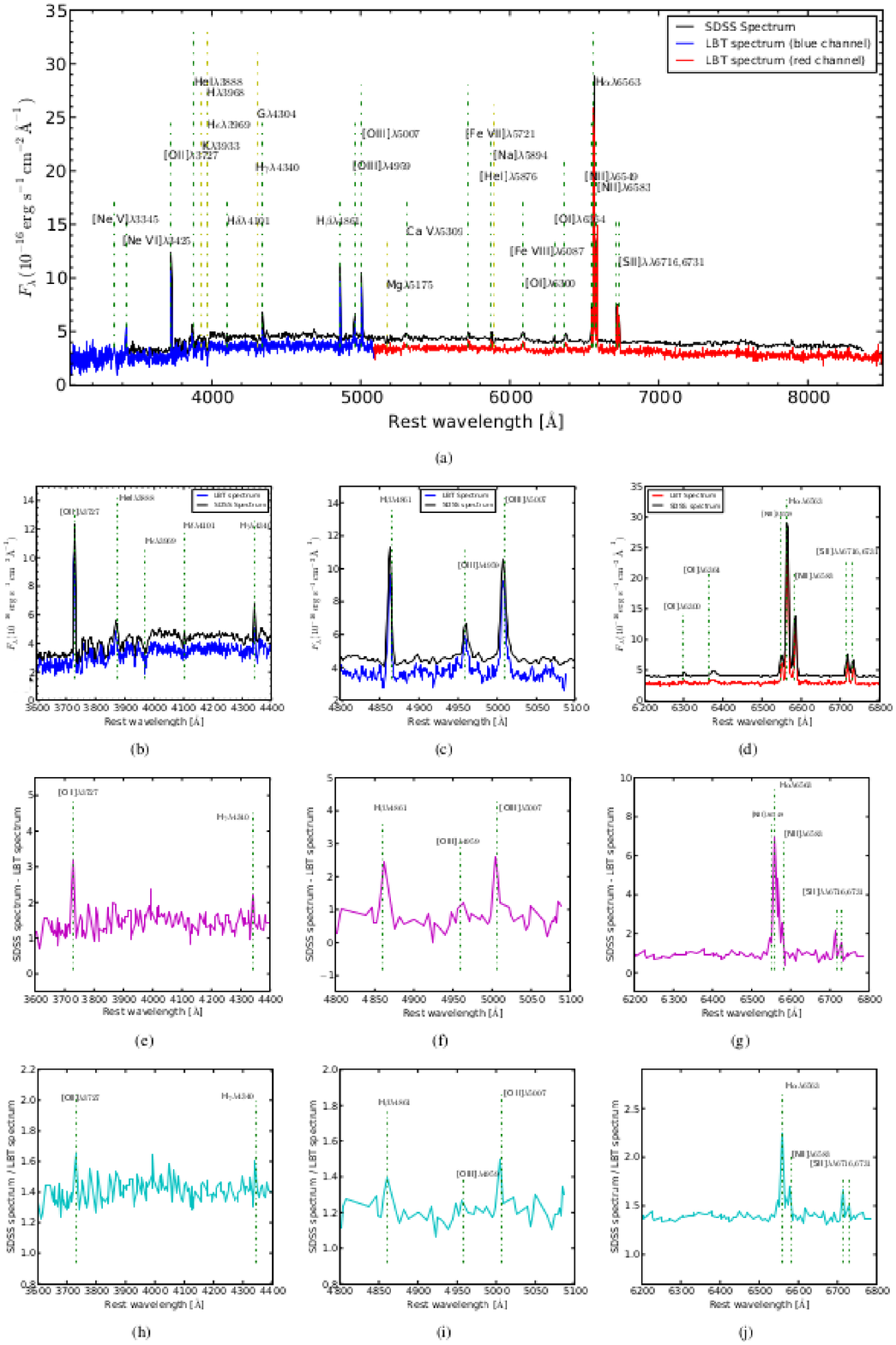}
\caption{
The plot shows the optical spectrum of J0938 as observed by SDSS and LBT-MODS. 
The second row shows zooms into the spectrum in different regions. The third row shows 
the differences between SDSS and LBT spectrum in these regions, while the fourth row shows the ratios of these spectra.
}
\label{spec_j0938}
\end{center}
\end{figure*}

As we show in Fig.~\ref{fig:subtraction_rest}, and as is evident from the
data in Tab.~\ref{all_sources2}, the host of the source is extended. 
Comparison with stars in the field show that 
about 50\% to 60\% of the continuum flux in the LBT 0.8'' slit is due to the 
unresolved nuclear component. The rest can be attributed to the a contribution of 
the extended host.
Hence the estimated continuum variability is probably only an upper limit.
However, for completeness we leave the variability estimates in Fig.~\ref{cont_vari}
as upper limits. 
The median and median deviation derived from this plot are not effected by this.
In Fig.~\ref{spec_j0938} we over-plot the two optical spectra
(LBT and SDSS)  of J0938.  The LBT spectrum was observed in 2012, and the
SDSS spectrum was taken in 2006. Therefore, we can
discuss the variability on a time scale of 6 years.
We found that, for this object, there are two types of variability: first,
there is variability in the continuum level between the two spectra, where
the continuum level of the SDSS spectrum is higher by a factor of
$\sim$1.3 than that of the LBT spectrum.
This can be seen in Fig.~\ref{spec_j0938} (a,b,c,and d) where we plot different sections of the
J0938 spectrum.
The second type of the variability is in the emission lines. To show this we subtract the 
two spectra from each other and plot the difference (Fig.~\ref{spec_j0938} e, f, and g). 
Additionally, we show the variability of these lines by displaying the ratio of the 
two spectra (Fig.~\ref{spec_j0938} h,i, and j). We found that the emissions lines 
that varied most in the spectrum of J0938 are [\ion{O}{ii}], $\mathrm{H\beta}$, 
[\ion{O}{iii}]$\lambda$5007, and  $\mathrm{H\alpha}$.
The $\mathrm{H\alpha}$ and [\ion{O}{iii]} lines show a variation of the order 1.25.
The measured line fluxes are listed in Tab.~\ref{fluxes_fwhm}.

\subsection{SDSS J120300.19+162443.8}
\label{sec:j1203}
SDSS J120300.19+162443.8 (henceforth J1203) has been mentioned in \cite{2012MNRAS.421.1043S} among 2865 galaxies 
to have a strong nebular He~II$\lambda$4686 emission.  A strong  He~II$\lambda$4686 line indicates that the
nuclear radiation field of these objects is dominated by highly ionizing radiation.
The ionization potential of $\mathrm{He^{+}}$ is 54.4 eV corresponding to a UV photon wavelength of $\lambda \approx 228 \AA$.

In  Tab.~\ref{fig:variab} we list the flux densities $\mathrm{{f}_{\lambda}}$ of J1203 at different wavelengths
obtained from our LBT data and spectroscopy and photometry data as listed by SDSS.
We note that the flux values of the continuum at different wavelength indicate variability. 
Moreover, the power law index of the continuum feature of both observations SDSS (photometric \& spectroscopy) 
and LBT (spectroscopy) has also varied as we show in the optical spectrum of J1203 
in Fig.~\ref{spec_j1203}.\\

The optical emission line spectrum of J1203 is dominated by a NLR (see Fig.~\ref{spec_j1203}).
In Tab.~\ref{fluxes_fwhm} we see that all the emission lines are narrow with FWHM values 
of $\leqslant 8.3 \AA$ (where $8.3 \AA\ \approx 500$ $\mathrm{km}\ \mathrm{s}^{-1}$).
J1203 has very low count rates at the continuum level.
The H$\alpha$ complex gives indications of a possible
faint broad component.
\cite{Barth2014} find that all S2 candidates in their sample for which
high spectral resolution deep exposures had been taken showed 
some broad line emission that was not visible in previous spectra.
A comparison between the 
[OIII] $\lambda$5007 and H$\beta$ line shows a very similar 
line profile, within the uncertainties, hence we classify J1203 as a S2 galaxy.

The over-plot of the two optical spectra (LBT and SDSS) 
for J1203 is presented in Fig.~\ref{spec_j1203}. The SDSS 
spectrum was observed in 2007, while the LBT 
spectrum was observed in 2012. Hence, for 
this galaxy we can discuss the variability that happened over 5 years. 
We find that the continuum  variability here is rather limited 
to the blue channel region from $3300\AA$ to $3700\AA$ where effects of red wing of
the "Big Blue Bump BBB" may well be present \citep{2008RMxAC..32....1G, 2001tysc.confE.198S}, 
as shown in Fig.~\ref{spec_j1203} (d). 
The observed continuum of the 
LBT spectrum in this region is higher than the SDSS spectrum. But 
for the rest of the spectrum both spectra are at the same level till 6000 \AA. 
The line emission varies by about 40\% (see Tab.~\ref{fluxes_fwhm}).

\subsection{SDSS J115816.72+132624.1}
\label{sec:j1158}
The optical spectrum of SDSS J115816.72+132624.1 
(hereafter J1158; observing dates see Tab.~\ref{all_sources1}) 
is strongly dominated by \ion{Fe}{ii} lines 
\citep[as can be seen in Fig.~\ref{spec_j1158} 
and as reported also in][]{2010ApJS..189...15K, 2011ApJ...736...86D}. 
Tab.~\ref{fig:variab} shows that the flux values of the photometric and spectroscopy observations 
obtained within the  SDSS survey are in good agreement. However, both values show a difference with respect to 
the flux value extracted from the LBT spectrum, indicating variability. 
In \cite{2009MNRAS.398..109W} the authors refer to J1158 as a faint source and mention it in the 
Imperial IRAS-FSC Redshift Catalogue (IIFSCz) among 60303 galaxies. Verifying the consistency of source positions 
they improved the optical, near-infrared, radio identification. 
According to \cite{2010ApJ...714..367Z} J1158 is classified as a low-ionization broad absorption-line (loBAL) 
quasar, with \ion{[Mg}{ii]} absorption-lines with a width of 
$\Delta \upsilon_{c} \geqslant 1600\ \mathrm{km}\ \mathrm{s}^{-1}$. 
Additionally, the authors calculate the continuum luminosity at 
$5100\AA$ ($\lambda L_{5100}$) as
$7.7 \times 10^{44}\mathrm{erg}\ \mathrm{s}^{-1}$. 
From the LBT spectrum we obtain a luminosity at the same wavelength with a flux of 
$2.81 \times 10^{45}\mathrm{erg}\ \mathrm{s}^{-1}$. The SDSS spectrum results in a value of 
$2.25 \times 10^{45}\mathrm{erg}\ \mathrm{s}^{-1}$. 
Consistent with the fluxes listed in Tab.~\ref{fig:variab} we can conclude that 
the continuum luminosity of this source has varied by a factor of 1.2 over the past 8 years.

In Fig.~\ref{spec_j1158} we present the over-plot of the LBT and SDSS spectra for J1158. The LBT data have been obtained in January 2012, and the SDSS survey monitored this source in May 2006. Hence, for this object we are able to study and discuss the spectral variability on a 
time scale of 6 years. After over-plotting both spectra and measuring the flux densities of the emission lines (Tab.~\ref{fluxes_fwhm}) 
for this galaxy, we found that the variability in the continuum level and the continuum spectral index (hereafter PLI)
may well be coupled to each other as the red tail of the BBB may be the dominant contributor to the variability behavior.
In addition we find strong variability in emission line intensity.

As shown in Fig.~\ref{spec_j1158} (a,b,c, and d) the continuum level of the LBT spectrum is higher 
than that in the SDSS spectrum by a factor of about 1.25.
This could be linked to power-law variability over the entire optical spectrum
as expected from localized temperature fluctuations of a simple inhomogeneous disk 
\citep{2014ApJ...783..105R}.
We find the PLI for both spectra to vary from
 -0.21 $\mathrm{erg}\ \mathrm{s}^{-1}\ \mathrm{cm}^{-2}$ / \AA ~~for the  SDSS spectrum 
to the PLI of LBT spectrum of -0.01 $\mathrm{erg}\ \mathrm{s}^{-1}\ \mathrm{cm}^{-2}$ / \AA. 
Among them the H$\beta$$\lambda 4861$ and \ion{[O}{iii]}$\lambda 5007$ line show the 
strongest variation by a factor of about 3.
In general, the emission lines in this source are more variable than in other objects of our sample.

\subsection{SDSS J091146.06+403501.0}
\label{sec:j0911}
SDSS J091146.06+403501.0 (hereafter J0911) was first reported in \cite{2002ApJS..143....1M} 
as a radio source. In the third data release of the SDSS survey \citep{2005AJ....130..367S} it is classified as a quasar. 
\cite{2010ApJ...714..367Z} classify J0911 as a low-ionization broad absorption-line (loBAL) quasar, with \ion{[Mg}{ii]} absorption-line in its 
spectrum with a line width of  $\Delta \upsilon_{c} \geqslant 1600\ \mathrm{km}\ \mathrm{s}^{-1}$. 
J0911 exhibits strong 
\ion{Fe}{ii} lines as shown in Fig.~\ref{spec_j0911}. 
Based on the SDSS DR5 data 
\cite{2010ApJ...714..367Z} calculate the continuum luminosity of J0911
at $5100\AA$ ($\lambda L_{5100}$) to be $2.9 \times 10^{44}\mathrm{erg}\ \mathrm{s}^{-1}$, while at
the same wavelength we derive from our LBT spectroscopy data a luminosity value of 
$3.66 \times 10^{44}\mathrm{erg}\ \mathrm{s}^{-1}$, and for the SDSS spectrum we find 
$2.84 \times 10^{44}\mathrm{erg}\ \mathrm{s}^{-1}$. 
These three values, and the
measurements in Tab.~\ref{fig:variab}, show that there is significant
variability in the continuum emission of this source over
these two epochs (see Fig.~\ref{spec_j0911}).
This is in contrast to the strong variability we find for J1158 and supports the consistency of the 
calibration between the different data sets.
As part of the SDSS survey this source has been observed in March 2003, 
while the LBT spectrum of the source was taken in February 2012. 
Hence we look at a time span of 9 years for the variability search.
The optical spectra of J0911 can be seen in Fig.~\ref{spec_j0911} (a, b, c, and d). 
As illustrated in this figure, the variability of the source is small compared to other sources in the
sample, except for the blue part of the spectrum in Fig.~\ref{spec_j0911} (b), 
where we see a difference in the continuum slope. 
This variation in the blue part of the continuum may caused by an accelerating 
outflow emanating from the black hole in the center of the galaxy, see \cite{2010A&A...509A.106S}.  
In Tab.~\ref{fluxes_fwhm} we list the flux measurements of the 
emission lines and their FWHM. 

\subsection{SDSS J080248.18+551328.9}
\label{sec:j0802}
SDSS J080248.19+551328.9 (in the following J0802) was first mentioned in the  
Fifth Data Release DR5 of the SDSS catalog for quasars \citep{2007AJ....134..102S}. 
The initial redshift determination
$\mathrm{z}= 0.66287\pm0.00107$ in DR5 \citep{2007AJ....134..102S},
later has been corrected to a value of 
$\mathrm{z}=0.664065\pm0.000355$ \citep{2010MNRAS.405.2302H}.
\cite{2010AJ....139.2360S} and \cite{2012AJ....143..119I} classified J0802 as a quasar, 
which is supported by \cite{2009ApJ...692..758G},
\cite{2009ApJ...698..819L} and
\cite{2011MNRAS.410..860A} with reference 
to a broad absorption line of \ion{[Mg}{ii]} $\lambda 2800$ in its spectrum.
From the optical spectra it is apparent that J0802 is highly dominated by \ion{Fe}{ii} lines (see Fig.~\ref{spec_j0802}).
We notice that the continuum power law index derived from the 
LBT and SDSS spectroscopy data as well as the value derived from SDSS photometry are in good agreement (Tab.~\ref{fig:variab}).
According to these data the continuum spectrum peaks in the i-band.
However, comparing the flux densities we find that while values of SDSS photometric and LBT spectroscopy are similar 
they both are different to what can be derived from  SDSS spectroscopy.
This indicates that J0802 shows some continuum variability between epochs.

The LBT spectrum was observed in January 2012, while the SDSS spectrum was 
 taken in January 2005, resulting in a time baseline of 7 years for this source.
Plotting both spectra (SDSS \& LBT) for J0802 in Fig.~\ref{spec_j0802} (a, b, c, and d) we 
find that there are spectral similarities to the source J0938.  
Both spectra have the same continuum slope but the 
continuum level of the LBT spectrum is higher by a factor of 1.3 compared to the SDSS spectrum.
The intensity of the emission lines in LBT spectrum are stronger 
than those obtained from the SDSS spectrum (see Tab.~\ref{fluxes_fwhm}). 
The H$\gamma$ $\lambda$4340 line shows the strongest variation with a factor of about 2.

\subsection{2MASX J035409.48+024930.7}
\label{sec:j0354}
This source 2MASX J035409.48+024930.7 (hereafter J0354) is one of the two sources that have not been covered by the SDSS survey. 
J0354 was serendipitously discovered and discussed, together with 18 other objects, in \cite{1981ApJ...243L...5C} as an AGN X-ray source.
In \cite{1982ApJ...257...40B} the authors report that the J0354 spectrum ($\mathrm{M_{V} \sim 18}$)
allows for either QSO or Seyfert 1 classification. They outline that compared to other samples \citep[e.g.,][]{1982ApJ...253L...7M}
the object is amongst the brightest X-ray sources.
They also report that this source has a large ratio of X-ray to optical luminosity $\mathrm{L}_{x}/\mathrm{L}_{opt}=2$
which is above the typical ratio of 0.5 found for quasars.
\cite{1991A&A...245..423H} state that J0354 is a gas rich galaxy that arose from a two-galaxy interaction, 
where the larger one is either a quasar or S1 and the second clearly shows signs of tidal distortion. 
 \cite{1982ApJ...261L..23H} suggest that the companion is bluer than the quasar. 
\cite{2006A&A...455..773V} classify the source as S1.5.\\
We can compare our LBT spectroscopy data for J0354 with spectroscopy data obtained in Oct 2004 
by the 2.4 m Hiltner Telescope at MDM Observatory at Kitt Peak, Arizona, USA published by \cite{2005AJ....130..355G}. 
Here the source was observed with 1.5'' and 2'' slits under moderate seeing conditions.
We notice that the continuum level of the LBT spectrum is different than the spectrum of Hiltner Telescope as shown 
in Fig.~\ref{spec_j0354}, where the continuum level of J0354 taken with the Hiltner telescope is higher 
than the LBT spectrum for the same source, and the emission lines of their observation are more intense 
than our observation. Additionally,  \cite{2005AJ....130..355G} report line fluxes of 
[[\ion{O}{iii}] $\lambda4959$ and $\lambda5007$ as 
$(130\pm6)$ and $(413\pm15) \times 10^{-16}\mathrm{erg}\ \mathrm{s}^{-1}\ \mathrm{cm}^{-2}$, respectively.
From the LBT data we obtain fluxes of the same lines as  
$(90)\times 10^{-16}\mathrm{erg}\ \mathrm{s}^{-1}\ \mathrm{cm}^{-2}$  and
$(505)\times 10^{-16}\mathrm{erg}\ \mathrm{s}^{-1}\ \mathrm{cm}^{-2}$,  respectively.
The comparison shows that the [\ion{O}{iii}] $\lambda5007$ line varies by a factor of almost 50\%.

\subsection{GALEXASC J015328.23+260938.5}
\label{sec:j0153}
GALEXASC J015328.23+260938.5 (henceforward J0153) is the second source that is not covered by the SDSS survey. 
This source is present in the Infrared Astronomy Satellite (IRAS) Point Source Catalogue 
\citep[][using the name IRAS 01506+2554]{1988iras....1.....B} and in the 2nd XMM-Newton Serendipitous Source Catalog 
\citep[][ using the name 2XMM J015328.4+26093]{2007A&A...465..759M, 2009ApJ...705..454N}. 
From the IRAS survey, we know the flux density at two wavelengths, 60 and 100 microns, to be 0.5777 \& 1.144 Jy, respectively. 
In addition to our LBT observations 
(see Fig.~\ref{spec_j0153})
a spectrum of J0153 has been obtained by \cite{2002ApJ...581...96Z} in February 2002 with the Zeiss universal spectrograph 
located at the Beijing Observatory 2.16 m telescope, with a slit width of 2\farcs5 and seeing disk of 2\farcs5. 
Moreover, the spectral resolution determined on the night sky is $5.2\AA$ FWHM. 
They also found that J0153 is a radio-loud quasar with very strong \ion{Fe}{ii} emission lines, for 
the far- and near-infrared luminosity, they found 
$\mathrm{{L}_{FIR}} \approx {10}^{12.7}\ \mathrm{L}_{\sun}$ and $\mathrm{{L}_{NIR}} \approx {10}^{12.5}\ \mathrm{L}_{\sun}$.
\\
J0153 is one of 30 galaxies that \cite{2011A&A...528A.124C} study to follow the galaxy evolution and especially the 
star formation efficiency (SFE) using CO lines. They find that the SFE of this source is $>$ 802 
($\mathrm{{L}_{\sun}}/\mathrm{{M}_{\sun}}$) which is 4.7 times higher than the local ultra-luminous 
infrared galaxies (ULIRGs) (170 $\mathrm{{L}_{\sun}}/\mathrm{{M}_{\sun}}$), indicating highly efficient star 
formation activity.

\subsection{SDSS J034740.18+010514.0}
\label{sec:j0347}
SDSS J034740.18+010514.0 (henceforth J0347) was mentioned first time in a study for high-luminosity 
sources \citep[][using the name IRAS 03450+0055]{1988ApJ...327L..41L} and classified as quasar. 
In \cite{1991ApJS...75..297H} J0347 was classified as S1 galaxy by \cite{2006A&A...455..773V} as S1.5. 
With $\mathrm{{L}_{FIR}} \approx {10}^{10.2}$
\cite{1988ApJ...327L..41L} and \cite{1989ApJ...340L...1L} 
find J0347 to have the lowest luminosity among other quasars in their sample, 
additionally they find that this source has an extremely red optical continuum. 
Since no SDSS spectroscopic observations are available for this object, we used 
spectroscopic data taken on 21 September 2003 with the CARELEC spectrograph attached to the 1.93m 
telescope of the Observatoire de Haute-Provence (OHP) with a $2\arcsec$ slit under $2.5\arcsec$ seeing.
\\
\cite{1996A&A...314..419G} search for variability in a sample of 12 narrow line Seyfert 1 galaxies, including J0347. 
They find 10 of these sources to be variable in the flux of the optical permitted lines over a period of one year 
and also found variation in the continuum level. 
The variability ratio they find in the permitted lines $\mathrm{H\beta}$ and $\mathrm{H\alpha}$ is 4.58 and 0.599 respectively.\\
We list measurements of the continuum flux from SDSS photometry as well as from 
OHP and LBT spectroscopy in Tab.~\ref{fig:variab}. 
Line flux measurements are presented in Tab.~\ref{fluxes_fwhm} (see also plots in Fig.~\ref{spec_j0347}).

\begin{table}
\begin{center}
\caption{Median and median deviations for the variability in the continuum \& emission lines.}
\begin{tabular}{c c c c c c c c c c c c c}
\hline
Type   &\multicolumn{2}{c}{Continuum}& \multicolumn{4}{c}{Narrow        } \\
       &\multicolumn{2}{c}{         }& \multicolumn{4}{c}{Emission lines} \\
       &$\mu$&$d\mu$&$\mu$&$d\mu$&\\
    \hline
BLS1\& QSO& 39.0&17.0&30.0&10.5\\
S2 \& NLS1& 27.0&9.0&15.0 & 6.0\\
\hline
\label{t-test1}
\end{tabular}
\end{center}
   \begin{flushleft}
 \it{\textbf{Notes:} $\mu$ are the median values, $d\mu$ the median deviations as defined in the text.}
      \end{flushleft}
\end{table}

\begin{table*}
\begin{center}
\caption{The T-test statistic for the variability in the continuum \& emission lines.}
\begin{tabular}{c c c c c c c c c c c}
\hline
Type   &\multicolumn{5}{c}{Continuum}& \multicolumn{5}{c}{Emission lines} \\
       &n &$\overline{\chi}$&$\sigma$&df&T-value&n &$\overline{\chi}$&$\sigma$&df&T-value\\
    \hline
BLS1\& QSO& 42& 42.25&20.76&\multirow{2}{*}{75}&\multirow{2}{*}{3.38}&18&29.48&10.71&\multirow{2}{*}{33}&\multirow{2}{*}{4.25}\\
S2 \& NLS1& 35& 29.51&11.72& &                     &17&15.51& 8.67&                   & \\
\hline
\label{t-test2}
\end{tabular}
\end{center}
\begin{flushleft}
 \it{\textbf{Notes:} Here $n$ denotes the number of sample elements,
$\overline{\chi}$ are the sample mean values,
$\sigma$ the sample standard deviations,
$df$ the degrees of freedom followed by the corresponding T-values.}
\end{flushleft}
\end{table*}

\begin{figure*}
\includegraphics[width=0.9\linewidth]{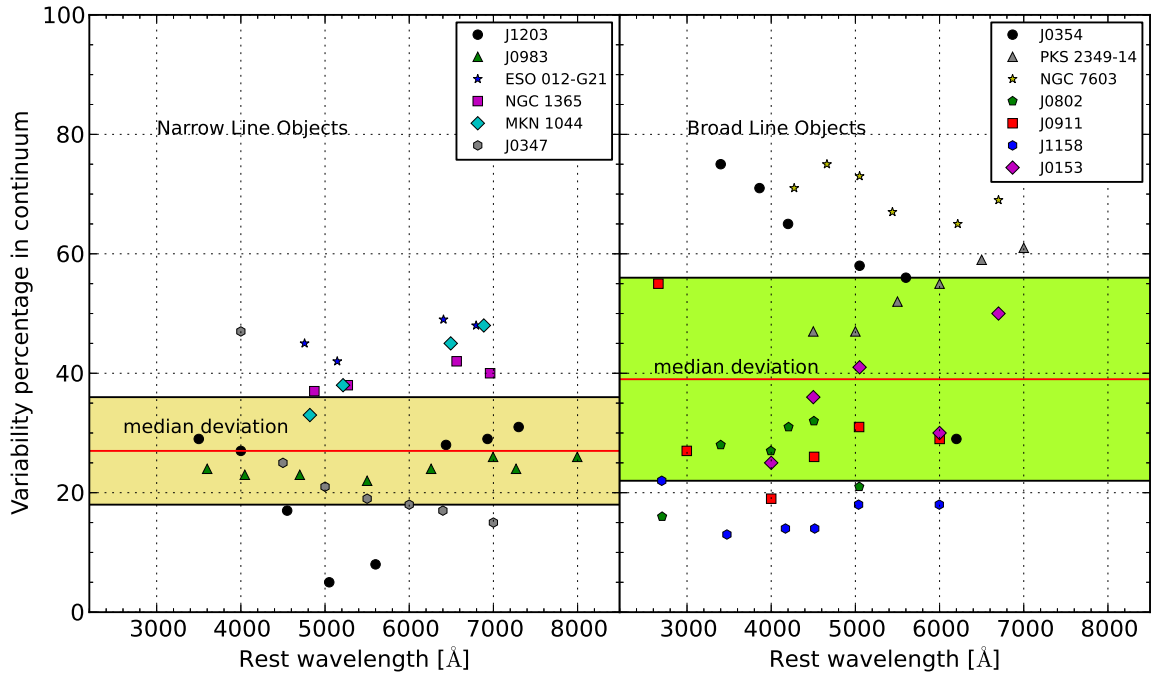}
\caption{
The variability percentage in the continuum. The data for the NLS1 sources 
ESO~012-G21, NGC~1365, and MKN~1044 are taken from Giannuzzo \& Stirpe (1996). The data for the BLS1 sources NGC~7603 and PKS~2349-14 
are based on spectra from Kollatschny et al. (2000) and Kollatschny et al. (2006a), respectively.
}
\label{cont_vari}
\end{figure*}

\begin{figure*}
\includegraphics[width=0.9\linewidth]{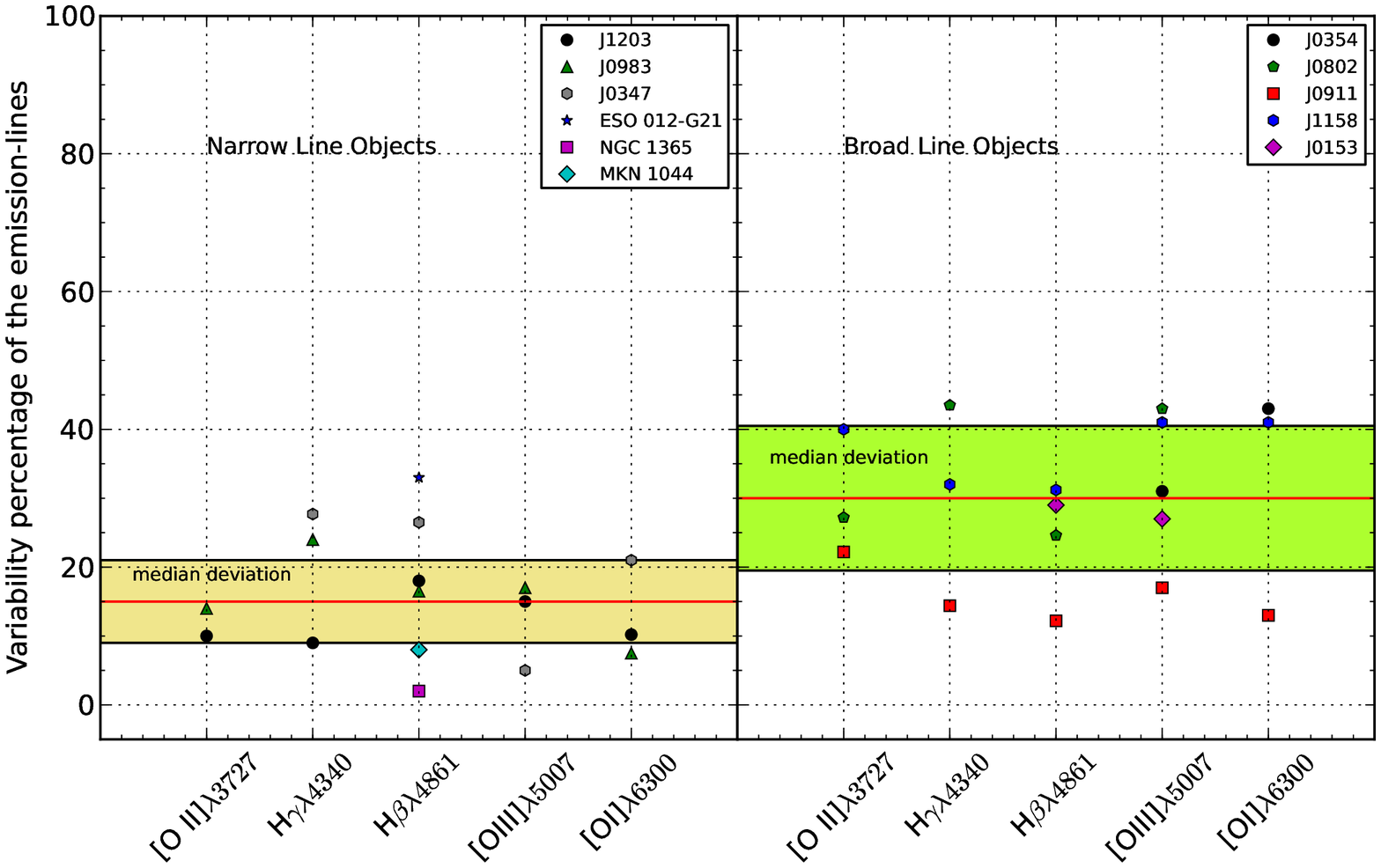}
\caption{
The variability percentage in the emission lines. 
The line measurements of ESO~012-G21, NGC~1365, and MKN~1044 are taken from Giannuzzo \& Stirpe (1996).
}
\label{lines_vari}
\end{figure*}

\section{Discussion}
\label{sec:discussion}
Our study allows us to compare the continuum and line variability of two different source samples
with different spectroscopic identifications that by themselves imply a different degree of activity.
We find that the Seyfert~2/NLS~1 sample shows in general a smaller degree of variability 
compared to the BLS1 and QSO sample.
In the following we probe the significance of this difference and present a possible physical
interpretation of the observed phenomenon.
We extend the data on the continuum variability with 
multi epoch information on ESO~012-G21, NGC~1365, and MKN~1044 from \cite{1996A&A...314..419G},
NGC~7603 from  \cite{2000A&A...361..901K} and  PKS~2349-14 from \cite{2006A&A...454..459K}.
We extended the data on multi epoch line variability with ESO~012-G21, NGC~1365 and MKN~1044 from \cite{1996A&A...314..419G}.

\subsection{Probing the degree of variability}
\label{subsec:variab}

Based on our detailed study of 18 sources 
(see Tab.~\ref{all_sources1}, Tab.~\ref{rev_maptab} and captions in Figs.~\ref{cont_vari} and \ref{lines_vari})
we find that there are significant differences in variability between 
S2\&NLS1 (8) sources and BLS1\&QSO (10) sources. 
This applies both for the continuum and the emission lines. 
In the following we discuss the relative variability $\Delta_{max}$ 
as the maximum difference $\delta_{max}$ with respect to the maximum value $S_{max}$,
i.e. $\Delta_{max}=\delta_{max}/S_{max}$ 
in (i.e. dependent on the inter-calibration uncertainties which are the same for both source classes).
Correction for relative measurement uncertainties $\sigma$ of the order of 10\% results in
\begin{equation}
\Delta_{max,corr.}=\sqrt{\Delta_{max}^2-\sigma^2}
\end{equation}

We find that the BLS1\&QSO have stronger variablity than S2\&NLS1. \cite{2013AJ....145...90A} came to the same conclusion: 
NLS1 have a systematically lower degree of variability 
if compared to BLS1 sources. This is in agreement with the existing anti-correlation between 
AGNs variability and Eddington ratio \cite{2011A&A...525A..37M, 2012ApJ...758..104Z}. 
Other authors \citep[e.g.,][]{2000ApJ...540..652W} concluded from variability studies that the blue 
part of AGN spectra is relatively stronger variable than the red part.
For our study we show this result in Fig.~\ref{cont_vari}. 
We notice that for J0354 and J0911 the variability in the blue part of the continuum is, 
larger than the variability in the red part.

To judge on the difference between the two source samples in their continuum and line variability
we calculated the median and median deviation as well as the mean and standard deviation.
We also chose the median since we do not know ab-initio if we have normal distributions.
Here, the  median deviation is defined as the median of the absolute differences between the median and the
sample values.
The median deviation is usually smaller than the standard deviation as it is less sensitive with respect to outliers.
From Tab.~\ref{t-test1} and Figs.~\ref{cont_vari} and \ref{lines_vari} we find that the median degree 
of continuum variability is 27\%$\pm$9\% for the S2/NLS1 and 39\%$\pm$17\% for the BLS1/QSO sources.
For the median line variability we find 15\%$\pm$6\% for the S2/NLS1 and 30\%$\pm$11\% for the BLS1/QSO sources. 
Furthermore, we find that the median values are separated by about one median deviation.

\begin{table}
\begin{center}
\caption{The variability percentage in continuum at rest wavelength 5100\AA 
and $\mathrm{H\beta}$, for AGNs taken from reverberation mapping (light curve).}   
\begin{tabular}{l c c c c}      
\hline
Sources &Cont. 5100\AA & $\mathrm{H\beta}$&Type  &References\\
        &\%            & \%               &  &\\
\hline 
Mrk~335 & 34 & 24 &NLS1&1,2\\
Mrk~6   & 36 & 25 &NLS1&1\\
3C~120  & 57 & 40 &BLS1&1\\
Ark~120 & 48 & 47 &BLS1&1\\  
Mrk~590 & 43 & 50 &BLS1&1\\ 
\hline                          
\end{tabular}
\label{rev_maptab} 
\begin{flushleft}
\it{\textbf{References.} (1)~\citet{1998ApJ...501...82P}, (2)~\citet{2012ApJ...755...60G}.}
     \end{flushleft}
\end{center}
\end{table}

\subsubsection{Robustness of the separation}
\label{subsec:variab-rob}
{\it Results for narrow and broad line components:}
In Fig.~\ref{lines_vari}
 we show the results for the combined narrow and broad line recombination line fluxes
in order to compare the data with literature values.
However, the difference in the variability behavior between the narrow line 
and the broad line sources (with 15\%$\pm$6\% and 30\%$\pm$11\%, respectively) 
also holds if we investigate the narrow line and broad line components we obtained from
fitting the hydrogen recombination lines (see Tab.~\ref{narr_bro}).
Here the median value for the variability of the narrow line objects  J0347, J1203 and J0938 of 20$\pm$4 lies well below the median
result for the broad line objects of 41$\pm$8 derived from the 4 objects J1158, J0911, J0002 and J0153.
Although the separation into the two different source classes has been done using an overall line
width criterion, we find that the difference is also present if we look at the NLR and BLR 
line components separately
(see discussion in section \ref{subsec:NLRresponse}).
\\
\\
{\it Statistical considerations:}
For the BLS1/QSO sample we see the tendency in some sources for the variability to be up to 
20\% stronger in the blue (3000\AA) compared to the red (close to 6000\AA). 
However, this tendency is not equally well fullfiled for all sources and
rather weak with respect to the overall variability range from about 20\% to 70\%.
However, here we are more interested in the overall continuum and line variability.
Both are estimated for different sources and in different lines or spectral regions
of the continuum, which are all subject to different wavelength and line strength dependant calibration uncertainties.
Hence, we assume that for the two source samples the variability estimates are sufficiently independent, and can
be represented by a mean distribution.
In this case we can apply a T-test \citep{ttest} to determine the probability that the data sets from both the continuum and line 
variation originate from different distributions. The relevant test quantities are given in Tab.~\ref{t-test2}.
\\
We applied a T-test for two sample experimental statistics for independent groups.
The T-value was calculated via

\begin{equation}
T=\frac{\overline{\chi_1}-\overline{\chi_2}}{\sigma_{diff}}~~.
\end{equation}

The standard error of the difference we calculated via

\begin{equation}
\sigma_{diff}=\sqrt{\sigma^2_{\overline{\chi_1}} + \sigma^2_{\overline{\chi_2}}}~~. 
\end{equation}

Since we do not know the population standard deviation we use the sample standard deviation to estimate the standard error:

\begin{equation}
\sigma_{\overline{\chi}}=\frac{\sigma}{\sqrt{n}}~~,
\end{equation}

$n$ being the number of sample elements.
The number of degrees of freedom $df$ as used for two independent groups we obtained as 

\begin{equation}
df = (n_1 - 1) + (n_2 - 1)~~~.
\end{equation}

The probabilities that describe the matching of the two samples are tabulated \citep{ttest} as a function 
of the T-values and the degrees of freedom $df$.
We find that the continuum and line variability of the BLS1 and QSO sample originated with a better than 99.9\% probability from a 
different distribution compared to the Seyfert~2 and NLS1 sample. 
\\
\\
{\it Comparison to other samples:}
The separation between the narrow line and the broad line sources is
best fulfilled for the line variation. 
As shown in Tab.~\ref{rev_maptab} our results are very much consistent with those 
for a set of NLS1 and BLS1 sources obtained
from reverberation measurements carried out by 
\cite{1998ApJ...501...82P} and~\cite{2012ApJ...755...60G}.
\\
The findings for our small sample are also in full agreement with the results of a comparative study 
of the optical/ultraviolet variability of 
NLS1 and BLS1-type sources by \cite{Ai2013} (based on 55 NLS1- and 108 BLS1-type nuclei).
This underlines the representative character of our sample.
The authors show that the majority of NLS1-type objects exhibit significant variability on timescales from about 
10 days up to a few years, but on average their variability amplitudes  are much smaller compared to BLS1-type sources
\citep[see also,][]{Yip2009, Barth2014}.
Hence, in summary, we assume that the difference in continuum and line variability between the two samples is 
sufficiently significant to search for a physical explanation for the phenomenon.

\subsection{Accretion dominated variability}
\label{subsec:accretion}

Here we investigate how the difference in narrow line variability between the objects with spectra dominated by narrow lines
and broad lines is linked to the continuum variations of the active nuclei in the different samples.

In the following we denote with $\Delta$ differentiation with respect to the different samples, i.e.
$\Delta = \frac{d}{d~~sample}$.
If the variability is due to mass accretion $\frac{d}{dt}M$ onto a super-massive 
black hole with mass $M$ we can write the continuum luminosity $L_{cont.}$ as

\begin{equation}
L_{cont.} \propto M \times \frac{d}{dt}M~~.
\end{equation}

The expected difference in continuum variability between the samples can then be expressed as

\begin{equation}
\Delta L_{cont.} \propto  \Delta M \times \frac{d}{dt}M +  M \times \Delta \frac{d}{dt}M~~~.
\label{eq:eee1}
\end{equation}

For recombination lines the variation of the nuclear continuum results in a variation of the 
number of atomic species that reach the next higher state of ionization out of which they can recombine. 
For collisionally excited states this next higher state of ionization is the state in which they
can be observed at low densities emitting forbidden narrow line radiation.
In the following we use the case of the recombination lines to discuss the effect of the
reverberation.
Since we are mainly interested in the variation of the narrow line luminosity $L_{line}$ as a reverberation response 
to the variations in continuum luminosity we can write:

\begin{equation}
L_{line} \propto \frac{h\nu_{line}}{\langle hv \rangle } \frac{\alpha_{eff}}{\alpha_B} \frac{d}{dt}M
= \xi \frac{d}{dt}M;
\end{equation}

Here $\nu_{line}$ is the line frequency, $\alpha_{eff}$ and $\alpha_B$ are the effective and Menzel's case $B$
specific recombination coefficients and $\langle hv \rangle$ denotes the mean energy per photon.
Through the latter the combined coefficient $\xi$ depends on the ration between the total illuminated cross-section $A_C$
and ionized volume $V_C$ of the reverberating clouds in a single galactic nucleus. 
This ratio results in a size $l$ of the reverberating volume which has a square root dependency 
on the continuum luminosity: 

\begin{equation}
\xi \propto \frac{V_C}{A_C} \propto l \propto \sqrt{L_{cont.}}~~~.
\end{equation}

Therefore, combining the above equations for the line luminosity we can write:

\begin{equation}
L_{line} \propto \sqrt{M} (\frac{d}{dt}M)^{\frac{3}{2}}~~~.
\end{equation}

The expected difference in line variability between the samples can then be expressed as

\begin{equation}
\Delta L_{line} \propto  \frac{1}{2} M^{-\frac{1}{2}} \Delta M (\frac{d}{dt}M)^{\frac{3}{2}} +
 M^{\frac{1}{2}} \frac{3}{2} (\frac{d}{dt}M)^{\frac{1}{2}} (\Delta \frac{d}{dt}M)~~~.
\label{eq:eee2}
\end{equation}

We now assume that - as a second order derivative - the variation 
$\Delta \frac{d}{dt}M$ 
of the accretion stream onto the super-massive  black holes from sample to sample is small 
and a finite difference $\Delta M$ in total mean black hole mass per sample needs to be considered.
In such a scenario, the
strength of the accretion stream as such is the 
dominant quantity for the variations in continuum and line luminosity between 
between different sample members and the two sources sample in general.
We can then neglect the second terms in equations 
\label{eq:eee1} and \label{eq:eee2}
and write for the difference in continuum and line luminosity:

\begin{equation}
\Delta L_{cont.} \propto  \Delta M \times \frac{d}{dt}M \propto \frac{d}{dt}M 
\end{equation}
\begin{equation}
\Delta L_{line} \propto  \Delta M (\frac{d}{dt}M)^{\frac{3}{2}} \propto (\frac{d}{dt}M)^{\frac{3}{2}}~~~.
\end{equation}

Hence the increase in activity in continuum and line emission is due to the difference in $\Delta M$ between the samples,
however, there is an additional modulation for the variability of the line emission that is 
a function of $\frac{d}{dt}M$.
Combining these results we find:

\begin{equation}
\Delta L_{line} \propto (\Delta L_{cont.})^{\frac{3}{2}}~~~. 
\label{equ:relation}
\end{equation}

This simple relation was obtained by assuming that the difference in accretion streams between the two samples is 
- to first order - negligible.
The accretion stream onto the super-massive black hole is probably closely tied to the properties of the
interstellar matter in its immediate vicinity. Hence, it may be different from source to source and may not strongly depend on
the mass of the central super-massive black hole. However, if $\Delta \frac{d}{dt}M$ needs to be considered it will have an influence on the 
general trend of the derived relation (i.e the overall slope and curvature of the trend). In fact, it may be responsible for part
of the scatter of data points about this relation. However, we assume the second order derivative to be of lesser importance than the 
first order derivatives. This is supported by the fact that the general trend of the distribution of data points 
in Fig.~\ref{varplot} is well described.

Looking at the median  and mean values for the different samples in 
Tabs.~\ref{t-test1} and \ref{t-test2} we find that this condition described by 
equation~\ref{equ:relation} is to 
first order fullfiled to within about 15\% : 
For the means of the continuum variability we find  $(39/30)^{\frac{3}{2}}=1.5$ whereas 
the measured ratio of the means of the line variability is $27/15=1.8$.
For the median values of the continuum variability we find  $(42/35)^{\frac{3}{2}}=1.3$ whereas the measured 
ratio of the median values of the line variability is $29/16=1.8$.
This proportionality relation between the continuum and narrow line variability is demonstrated in Fig.~\ref{varplot}
and discussed in the following section.

\subsection{NLR response to the nuclear continuum variability}
\label{subsec:NLRresponse}

\begin{figure}
\includegraphics[width=0.9\linewidth]{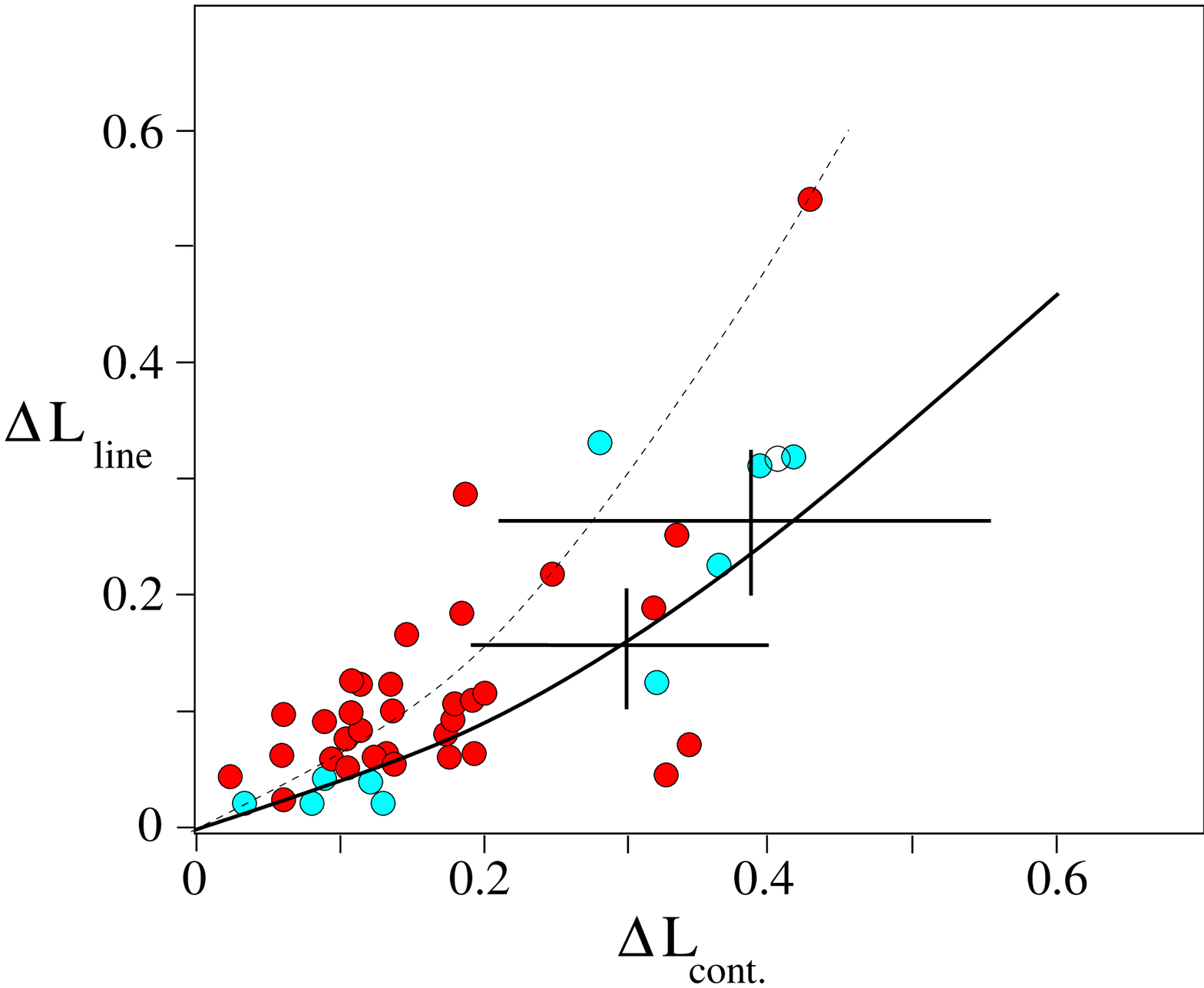}
\caption{
The continuum variability plotted against the line variability.
The solid and dashed black lines represent the relation 
$\Delta L_{line} = \epsilon (\Delta L_{cont.})^{\frac{3}{2}}$
(see text). Here $\Delta L$ indicates the observed luminosity variation
in line and continuum. The filled colored dots are QSO variability data (H$\beta$ and continuum) from 
\citet{Kollatschny2006}, \citet{Peterson2004}, and \citet{Kaspi2000}.
The Black crosses represent the median values and their uncertainties as 
we derived them for our sample (H$\beta$ and other lines plus continuum) and
as they are shown in Figs.~\ref{cont_vari} \& \ref{lines_vari}.
}
\label{varplot}
\end{figure}

In Fig.~\ref{varplot} the black crosses represent the median values and their uncertainties as 
shown in Figs.~\ref{cont_vari} \& \ref{lines_vari}.
In this plot the curved lines represent the relation 
$\Delta L_{line} = \epsilon (\Delta L_{cont.})^{\frac{3}{2}}$,
with $\epsilon$=1.0 for the thick straight line which follows
our data surprisingly well.
There are two main contributers $\eta_{star}$ and $\eta_{fill}$ to that value such that
$\epsilon=\frac{\eta_{star}}{\eta_{fill}}$.
For our sample, of the order of 50\% of the continuum flux density
is likely to be due to the stellar continuum, which is not variable on the time scales discussed here.
For more luminous objects this may be of the order of 20\% \citep{Kollatschny2006}.
This indicates that the pure nuclear continuum variability which is not contaminated by stellar flux 
may in general be larger by up to 50\%. This results in a value of $\eta_{star} \sim$1.9 
(dashed line in Fig.~\ref{varplot}).
However, the covering factor $\eta_{fill}$ of the clouds that are illuminated by the 
nucleus can be of the same order of magnitude.
Hence, with a considerable scatter, $\epsilon$ may in fact be close to unity,
such that the bold black line in Fig.~\ref{varplot} that
matches the observed data is an acceptable representation of the 
relation between the continuum and narrow line variability.

We have carried out the interpretation of the line response using the reverberation formalism.
This is usually applied to broad lines only and used to derive the size of the BLR,
making use of well sampled continuous light curves of the continuum and broad line emission, 
which are then cross-correlated.
In Fig.~\ref{varplot} we compare our finding with variability data from different broad line QSO samples.
The blue filled dots are QSO variability data from the sample listed in Tab.6 by 
\citet{Kollatschny2006}
the red filled dots are data from 
\citet{Peterson2004} and
\citet{Kaspi2000} as listed in Tab.8 by
\citet{Kollatschny2006}.
The comparison shows that the relation also holds for reverberation response of the broad line QSOs.
With the additional literature data Fig.~\ref{varplot} now comprises a total of 61 sources.
In the work we present here, however, the line response that we monitored in our small sample is dominated by 
forbidden narrow line emission.
Only the H$\beta$ line allows a separation between the broad and narrow line contribution.
However, the median and mean derived from the BLR component only is in good agreement with the overall result.
We base our investigation on typically 2 measurements over a baseline in time of the order of 10 years
(see Tab.~\ref{narr_bro}).
Reverberation response of the NLR based on measurements over 5-8 years has also been reported
for the radio galaxy 3C390.3 \citep{ClavelWamsteker1987} 
and for the S1 galaxy NGC~5548 \citep{Peterson2013}.
Within the SDSS survey one can also find example of multi-epoch observations that are consistent with this picture. 
In Fig.~\ref{SDSSnew} we show three epoch SDSS spectra for the NLS1 sources J014412 and J022205.
The variability estimates places them well into Fig.~\ref{varplot}.

Variability on time scale of years already sets an upper limit to the size of the corresponding region.
Since the variability is rather significant, the line flux contribution of that region is also high.
While there is a density and temperature gradient towards the nuclear position
\citep{2000Ogle, Penston1990, Mose1995} one also finds that the
surface brightness distribution of the NLR flux is very much centrally peaked.
The size of the NRL is luminosity dependant\citep{2003Schmitt}, however, 
the inspection of high angular resolution HST scans across S1 and S2 nuclei
shows that the unresolved central peak may contain 50\% or more of the
more extended nuclear forbidden line flux \citep{Bennert2006a, Bennert2006b}.
\citet{Peterson2013} find for NGC~5548 that 94\% of the NLR emission arises from within 18pc (54 ly).
\citet{2000Ogle} derive a density profile of r$^{-2}$ and 
\citet{Walsh2008} find that the [S II] line ratio indicates a radial stratification in gas density, with a sharp increase within
the inner 10-20 pc, in the majority of the Type 1 (broad-lined) objects.

Hence, the main response to the continuum variation must therefore originate in a rather compact region 
with a diameter of the order of 10 light years. 
This is in good agreement with the finding for 3C390.3 \citep{ClavelWamsteker1987} and NGC~5548 \citep{Peterson2013}.
Since this formalism fully explains the reverberation of the line emission with respect to the
continuum variation (with a surprisingly close proportionality; see Fig.~\ref{varplot}),
this suggests that the region illuminated by
the nucleus is almost fully spatially overlapping with the responding line emitting region. 
Hence, at larger distance the NLRs then show a much weaker response to the continuum variability 
most likely due to a combination of a lower overall luminosity and a lower volume filling factor.
The close correlation to the non-stellar nuclear continuum variability also implies that for the sources 
we investigated there are no other significant sources 
(i.e. star formation, extra nuclear shocks)
but the nuclear radiation field that contributes substantially to the line emission of the reverberating region.

\section{Conclusions}
\label{sec:conclusion}
We investigated a sample of 18 sources. 
For 8 objects in
Tab.~\ref{all_sources1}
we performed a detailed 
analysis spectroscopy over the available optical wavelength range. 
For 10 sources we obtained suitable multi epoch line and continuum data from the literature
(see Tab.~\ref{rev_maptab} and Figs.~\ref{cont_vari} and \ref{lines_vari})
covering time scales from 5 to 10 years.
Fig.~\ref{varplot} demonstrates that our findings describe the variability 
characteristics of a total of 61 sources.
From the line and continuum  variability of these active galactic nuclei we find the
following consistent picture that explains the differences between
the Seyfert~2/NLS~1 and the BLS1 and QSO sample:
The line luminosity $L_{line}$ can  be described as a reverberation response to the continuum luminosity $L_{cont}$.
In good agreement results of previous studies (references in section~\ref{subsec:NLRresponse})
we find that the bulk of the NLR emission arises from within 10-20 light years and shows a
reverberation response to variability of the nucleus.
The differences in variability do not require a variation of the accretion rate 
$\Delta \frac{d}{dt}M$ between the two samples. 
It is mainly the difference in black hole mass 
$\Delta M$ that is responsible for the difference between the samples. 
The increased variability of the line emission 
can fully be explained by the dependency on the accretion rate $\frac{d}{dt}M$.

\section*{Acknowledgments}
 Y.E. Rashed is supported by the German Academic Exchange Service (DAAD)
and by the Iraqi ministry of higher education and scientific research.
Y.E.R. wants to thank Dr. M.-P. V{\'e}ron-Cetty for
providing  us with an optical spectra for one of the source.
Macarena Garc\'{\i}a-Mar\'{\i}n is supported by the German
federal department for education and research (BMBF) under
the project number 50OS1101.
Gerold Busch is member of the \emph{Bonn-Cologne Graduate School of Physics and Astronomy}.

The spectroscopic observations reported here were obtained at the LBT
Observatory, a joint facility of the Smithsonian Institution and the University of Arizona.
LBT observations were obtained as part of the Rat Deutscher Sternwarten
guaranteed time on Steward Observatory facilities through
the LBTB cooperation.
This paper uses data taken with the MODS spectrographs built with funding from
NSF grant AST-9987045 and the NSF Telescope System Instrumentation Program (TSIP),
with additional funds from the Ohio Board of Regents and the Ohio State University Office of Research. 
The Sloan Digital Sky Survey is a joint
project of the University of Chicago, Fermilab, the Institute for Advanced
Study, the Japan Participation Group, Johns Hopkins University, the Max-Planck-Institute 
for Astronomy, the Max-Planck-Institute for Astrophysics, New Mexico
State University, Princeton University, the United States Naval Observatory, and
the University of Washington. Apache Point Observatory, site of the SDSS, is
operated by the Astrophysical Research C
consortium. Funding for the project has
been provided by the Alfred P. Sloan Foundation, the SDSS member institutions,
NASA, the NSF, the Department of Energy, the Japanese Monbukagakusho, and
the Max-Planck Society. The SDSS Web site is
http://www.sdss.org.

\bibliographystyle{mn2e}
\bibliography{Yasir_E_Rashed}

\begin{table*}
\begin{center}
\caption{\label{fluxes_fwhm} The line fitting results for J0938, J1203, 1158, J0911, J0802, J0354, J0153, and J0347.}
\begin{tabular}{l l r r r r r}
\hline
Sources&Em. lines                    & Ob. Wa.&\multicolumn{2}{c}{Inst.1}                             & \multicolumn{2}{c}{Inst.2}\\
\hline
       &                                  & &Flux                                &FWHM &Flux                                &FWHM\\
       &                                  & \AA &[$10^{-16}\ \mathrm{erg}$           & \AA &[$10^{-16}\ \mathrm{erg}$            &\AA \\
       &                                  & &$\mathrm{s}^{-1} \ \mathrm{cm}^{-2}$]& &$\mathrm{s}^{-1} \ \mathrm{cm}^{-2}$]            &    \\
\hline
      &&&&&&\\  
       &                                  & &\multicolumn{2}{c}{SDSS$^a$} & \multicolumn{2}{c}{LBT}\\
J0938  &   [\ion{O}{ii}] $\lambda$3727    &$ 4102\pm1 $&$   50.12\pm2.39  $&$ 5.66\pm0.49 $&$43.08  \pm2.38 $ & $6.05 \pm0.47 $\\
       &   {[\ion{Ne}{iii}]} $\lambda$3868&$ 4258\pm1 $&$   11.28\pm0.91  $&$10.67\pm0.93 $&$7.97   \pm1.14 $ & $8.38 \pm0.79 $\\
       &    H$\gamma$ $\lambda$ 4340      &$ 4777\pm1 $&$   25.18\pm1.58  $&$ 6.31\pm0.88 $&$18.95  \pm1.10 $ & $6.91 \pm0.71 $\\
       &   He II $\lambda$4686            &$ 5157\pm2 $&$   8.50 \pm0.73  $&$16.81\pm1.12 $&$4.04   \pm0.36 $ & $10.66\pm0.87 $\\
       &  H$\beta^*$ $\lambda$4861        &$ 5350\pm1 $&$   50.28\pm2.54  $&$ 5.61\pm0.49 $&$41.95  \pm2.83 $ & $5.91 \pm0.73 $\\
       &  {[\ion{O}{iii}]} $\lambda$5007  &$ 5510\pm1 $&$   45.06\pm1.57  $&$ 7.34\pm0.57 $&$37.43  \pm2.01 $ & $7.27 \pm0.66 $\\
       &  {[\ion{Fe}{vii}]} $\lambda$5721 &$ 6298\pm1 $&$   4.60 \pm0.39  $&$10.56\pm0.35 $&$6.55   \pm1.11 $ & $11.74\pm0.64 $\\
       &  {[\ion{Fe}{vii}]} $\lambda$6087 &$ 6699\pm1 $&$   11.55\pm0.75  $&$18.63\pm0.51 $&$11.19  \pm1.48 $ & $19.65\pm0.72 $\\
       & {[\ion{O}{i}]} $\lambda$6300     &$ 6935\pm1 $&$   4.94 \pm0.40  $&$ 6.75\pm0.41 $&$5.30   \pm1.14 $ & $9.14 \pm0.83 $\\
       & H$\alpha$ $\lambda$6563          &$ 7224\pm1 $&$  208.83\pm5.89  $&$ 7.75\pm0.57 $&$164.56 \pm5.03 $ & $6.43 \pm0.61 $\\
       & {[\ion{N}{ii}]} $\lambda$6583    &$ 7246\pm1 $&$   80.04\pm2.57  $&$ 7.77\pm0.52 $&$57.33  \pm2.70 $ & $6.17 \pm0.65 $\\
       &{[\ion{S}{ii}]} $\lambda$6716     &$ 7393\pm1 $&$   28.90\pm1.25  $&$ 7.46\pm0.54 $&$23.08  \pm1.45 $ & $5.92 \pm0.52 $\\
       &{[\ion{S}{ii}]} $\lambda$6731     &$ 7409\pm1 $&$   19.28\pm0.96  $&$ 7.15\pm0.56 $&$16.48  \pm1.32 $ & $6.02 \pm0.51 $\\
 \hline
&&&&&&\\
          &   & &\multicolumn{2}{c}{SDSS} & \multicolumn{2}{c}{LBT}\\
J1203&{[\ion{Ne}{v}]} $\lambda$3345       &$ 3898\pm1$ &$  20.42 \pm1.41 $ &$ 5.75\pm0.22 $&$27.82  \pm1.02 $ &$5.18 \pm0.31$\\
    &{[\ion{Ne}{vi}]} $\lambda$3425       &$ 3993\pm1$ &$  55.54 \pm2.84 $ &$ 4.67\pm0.25 $&$67.57  \pm1.93 $ &$5.08 \pm0.24$\\
    &{[\ion{O}{ii}] } $\lambda$3727       &$ 4344\pm1$ &$  83.03 \pm1.53 $ &$ 6.14\pm0.14 $&$93.17  \pm2.87 $ &$6.19 \pm0.22$\\
 & H$\epsilon$ $\lambda$3969              &$ 4625\pm2$ &$  38.36 \pm1.21 $ &$ 4.92\pm0.36 $&$48.01  \pm1.75 $ &$4.98 \pm0.76$\\
    & H$\gamma$ $\lambda$4340             &$ 5059\pm1$ &$  89.24 \pm1.49 $ &$ 5.64\pm0.41 $&$81.84  \pm2.04 $ &$5.41 \pm0.75$\\
     & H$\beta^*$ $\lambda$4861           &$ 5665\pm2$ &$  154.54\pm1.78 $ &$ 5.88\pm0.75 $&$126.39 \pm1.11 $ &$5.64 \pm0.97$\\
    &{[\ion{O}{iii}]} $\lambda$5007       &$ 5835\pm2$ &$  803.60\pm6.39 $ &$ 5.67\pm0.88 $&$681.90 \pm7.66 $ &$5.29 \pm0.85$\\
    &{[\ion{O}{i}]} $\lambda$6300         &$ 7343\pm3$ &$  12.53 \pm0.75 $ &$ 7.33\pm0.13 $&$11.25  \pm0.54 $ &$7.94 \pm0.12$\\
    & H$\alpha$ $\lambda$6543             &$ 7648\pm2$ &$  364.01\pm3.60 $ &$ 7.64\pm0.55 $&$168.34 \pm3.95 $ &$7.12 \pm0.71$\\
    &{[\ion{N}{ii}]} $\lambda$6583        &$ 7671\pm1$ &$  25.11 \pm1.63 $ &$ 6.59\pm0.18 $&$13.17  \pm0.94 $ &$6.31 \pm0.34$\\
    &{[\ion{S}{ii}]} $\lambda$6716        &$ 7827\pm3$ &$  17.09 \pm1.71 $ &$ 7.00\pm0.11 $&$10.37  \pm0.91 $ &$7.40 \pm0.14$\\
    &{[\ion{S}{ii}]} $\lambda$6731        &$ 7844\pm3$ &$  16.92 \pm1.28 $ &$ 7.81\pm0.12 $&$8.98   \pm0.88 $ &$6.61 \pm0.11$\\
   \hline
&&&&&&\\
            &  & &\multicolumn{2}{c}{SDSS} & \multicolumn{2}{c}{LBT}\\
 J0911&  {[\ion{Mg}{ii}]} $\lambda$2799   &$ 4030\pm1 $&$60.43  \pm1.53 $ & $29.01\pm0.95$&$   35.40\pm3.01 $&$29.01\pm1.32 $\\
 &  {[\ion{O}{ii}]} $\lambda$3727         &$ 5372\pm2 $&$11.80  \pm0.84 $ & $ 7.35\pm0.48$&$   15.26\pm0.55 $&$7.77 \pm0.32 $\\
 &  H$\gamma$ $\lambda$4340  (broad)      &$ 6258\pm2 $&$15.71  \pm1.14 $ & $54.14\pm1.87$&$   18.42\pm0.77 $&$51.78\pm2.01 $\\
  &  H$\gamma$ $\lambda$4340 (narrow)     &$ 6258\pm2 $&$6.92   \pm0.75 $ & $10.35\pm0.69$&$   8.03 \pm0.41 $&$11.06\pm0.73 $\\
 &  H$\beta$ $\lambda$4861 (broad)        &$ 7007\pm2 $&$46.05  \pm2.94 $ & $54.14\pm1.87$&$   52.55\pm2.49 $&$51.78\pm2.01 $\\
 &  H$\beta$ $\lambda$4861 (narrow)       &$ 7007\pm2 $&$17.09  \pm1.06 $ & $9.65 \pm0.62$&$  19.24 \pm0.82 $&$10.59\pm0.99 $\\
 &  {[\ion{O}{iii}]} $\lambda$5007        &$ 7218\pm1 $&$28.76  \pm1.53 $ & $9.88 \pm0.84$&$   34.49\pm1.55 $&$10.12\pm1.02 $\\
 &  {[\ion{O}{i}]} $\lambda$6300          &$ 9082\pm2 $&$4.75   \pm0.35 $ & $20.22\pm1.22$&$   5.43 \pm0.87 $&$16.14\pm1.76 $\\
\hline
&&&&&&\\
           & & &\multicolumn{2}{c}{SDSS} & \multicolumn{2}{c}{LBT}\\
 J1158 &{[\ion{O}{ii}]} $\lambda$3727     &$ 5366\pm1 $&$  10.61 \pm0.67  $&$8.35 \pm0.51$ &$7.12   \pm0.71 $ &$9.85 \pm0.88$\\
  & H$\gamma$ $\lambda$4340 (broad)       &$ 6245\pm3 $&$  58.82 \pm2.83  $&$75.32\pm2.92$ &$29.63  \pm2.16 $ &$94.16\pm2.56$\\
  & H$\gamma$ $\lambda$4340 (narrow)      &$ 6245\pm3 $&$  12.61 \pm0.56  $&$14.83\pm0.81$ &$7.40   \pm0.61 $ &$14.35\pm1.47$\\
  & H$\beta$ $\lambda$4861 (broad)        &$ 6994\pm3 $&$  201.01\pm3.31  $&$75.32\pm2.92$ &$105.69 \pm4.46 $ &$94.16\pm2.56$\\
  &  H$\beta$ $\lambda$4861(narrow)       &$ 6994\pm3 $&$  28.21 \pm0.96  $&$12.91\pm0.88$ &$17.05  \pm1.14 $ &$13.98\pm0.94$\\
  & {[\ion{O}{iii}]} $\lambda$5007        &$ 7208\pm3 $&$  59.41 \pm1.24  $&$12.01\pm0.91$ &$41.92  \pm1.88 $ &$13.65\pm1.18$\\
  & {[\ion{O}{i}]} $\lambda$6300          &$ 9071\pm3 $&$  5.81  \pm0.90  $&$6.55 \pm0.75$ &$3.98   \pm0.60 $ &$8.49 \pm0.98$\\
\hline
&&&&&&\\
          & & &\multicolumn{2}{c}{SDSS} & \multicolumn{2}{c}{LBT}\\
J0802   &{[\ion{O}{ii}]} $\lambda$3727   &$ 6203\pm1 $&$18.43  \pm1.23 $ & $15.48\pm1.31$&$  25.33 \pm0.99 $ &$12.91\pm0.89 $\\
   &H$\gamma$ $\lambda$ 4340 (broad)     &$ 7227\pm2 $&$33.82  \pm2.43 $ & $80.11\pm2.02$&$  60.36 \pm0.95 $ &$82.39\pm2.89 $\\
   &H$\gamma$ $\lambda$ 4340 (narrow)    &$ 7229\pm2 $&$8.62   \pm1.29 $ & $12.00\pm0.79$&$  15.41 \pm0.55 $ &$11.29\pm0.47 $\\
   &H$\beta$ $\lambda$4861 (broad)       &$ 8092\pm1 $&$84.23  \pm2.18 $ & $80.11\pm2.02$&$  95.07 \pm2.76 $ &$82.39\pm2.89 $\\
   &H$\beta$ $\lambda$4861 (narrow)      &$ 8092\pm1 $&$20.97  \pm1.63 $ & $11.98\pm0.85$&$  43.22 \pm1.54 $ &$13.65\pm1.98 $\\
   &{[\ion{O}{iii}]} $\lambda$5007       &$ 8331\pm1 $&$16.40  \pm1.42 $ & $11.51\pm0.81$&$  28.41 \pm1.62 $ &$13.41\pm1.12 $\\
     \hline   
     &&&&&&\\
 && &\multicolumn {2}{c}{Beijing$^b$(2)} & \multicolumn{2}{c}{LBT}\\
 J0153 &H$\beta$ $\lambda$4861 (broad)   &$ 6453\pm3 $ & $420.80  \pm-- $   & $ 77.01\pm-- $&$299.92 \pm10.98 $& $51.36\pm3.98$ \\
   &H$\beta$ $\lambda$4861 (narrow)      &$ 6453\pm3 $ & $99.60   \pm-- $   & $ 18.34\pm-- $&$68.20  \pm3.01 $& $14.96\pm1.34$\\
   &{[\ion{O}{iii}]} $\lambda$5007       &$ 6654\pm1 $ & $155.90  \pm-- $   & $ 13.37\pm-- $&$113.36  \pm4.51 $& $14.12\pm2.45$\\
\hline
\end{tabular}
 \begin{flushright}
 \it{ Continued on next page} \\
      \end{flushright}
\end{center}
\end{table*}
\setcounter{table}{8}
 \begin{table*}
 \begin{center}
\caption{{ Continued:} The line fitting results for J0938, J1203, 1158, J0911, J0802, J0354, J0153, and J0347.}
\begin{tabular}{l l r r r r r}
\hline
     &&&&&&\\    
   && &\multicolumn{2}{c}{Hiltner$^c$(1)} & \multicolumn{2}{c}{LBT}\\
 J0354  &{[\ion{Ne}{v}]} $\lambda$3346   &$ 3544\pm2 $ & $26.00   \pm10.00$ &$ 4.51\pm1.75 $&$11.91  \pm0.97 $& $3.09 \pm0.68$\\
   &{[\ion{O}{iii}]} $\lambda$5007       &$ 5183\pm1 $ & $413.00  \pm15.00$ &$ 5.51\pm0.94 $&$284.84 \pm4.42 $& $4.53 \pm0.26$\\
   &{[\ion{O}{i}]} $\lambda$6300         &$ 6522\pm1 $ & $15.15  \pm1.34  $ &$ 5.93\pm0.92 $&$8.70    \pm2.00 $& $6.60 \pm0.67$\\
   &{[\ion{S}{ii}]} $\lambda$6716        &$ 6952\pm1 $ & $14.45  \pm1.16  $ &$ 5.82\pm0.34 $&$ 10.90  \pm1.00 $& $5.79 \pm0.49$\\
   &{[\ion{S}{ii}]} $\lambda$6731        &$ 6967\pm1 $ & $12.52  \pm0.96  $ &$ 5.85\pm0.36 $&$ 8.40   \pm1.00 $& $5.09 \pm0.42$\\
\hline
      &&&&&&\\
      J0347&    & &\multicolumn{2}{c}{OHP$^d$} & \multicolumn{2}{c}{LBT}\\
   &H$\gamma$ $\lambda$ 4340  (broad)    &$ 4475\pm2 $ & $ 752.63 \pm18.28$ &$ 51.78\pm2.68$&$1094.56\pm9.19 $&$51.78\pm2.95$\\
   &H$\gamma$ $\lambda$ 4340 (narrow)    &$ 4475\pm2 $ & $ 273.21 \pm5.23 $ &$ 10.96\pm1.23$&$324.91 \pm9.19 $&$10.01\pm0.55$\\
   &H$\beta$ $\lambda$4861 (broad)       &$ 5010\pm1 $ & $1644.70 \pm16.46$ &$ 51.78\pm3.54$&$2280.33\pm21.87$&$51.78\pm0.92$\\
   &H$\beta$ $\lambda$4861(narrow)       &$ 5010\pm1 $ & $1058.20 \pm11.89$ &$ 13.18\pm1.02$&$1398.38\pm20.46$&$11.77\pm0.92$\\
   &{[\ion{O}{iii}]} $\lambda$5007       &$ 5156\pm1 $ & $951.10  \pm10.87$ &$ 12.71\pm2.33$&$996.74 \pm20.80$&$11.53\pm0.64$\\
   & {[\ion{O}{i}]} $\lambda$6300        &$ 6495\pm2 $ & $124.29  \pm2.67 $ &$ 16.19\pm1.11$&$157.30 \pm5.79 $&$18.78\pm1.78$\\
\hline
\end{tabular}
\begin{flushleft}
\it{\textbf{References.} (1)~\citet{2005AJ....130..355G}, (2)~\citet{2002ApJ...581...96Z}.\\
\textbf{Notes:}\\
($^*$) Means H$\beta$ are narrow.\\
($^a$) The Sloan Digital Sky Survey.\\
($^b$)  Beijing Observatory.\\
($^c$) Hiltner Telescope.\\
($^d$)  Telescope of Observatoire de Haute-Provence (OHP).
}
     \end{flushleft}
 \end{center}
\end{table*}


\begin{appendix}
\section \newline


\begin{table*}
\begin{center}
\caption{Broad and narrow emission lines variability.}
\begin{tabular}{c c c c c c}
\hline
Type                     & Sources &\multicolumn{2}{c}{Narrow line }          & \multicolumn{2}{c}{Broad line}           \\
                         & &\multicolumn{2}{c}{components   }          & \multicolumn{2}{c}{components  }           \\
                         &            &  line type & Varib. \%     & line type & Varib. \%     \\
\hline
 Narrow lines objects    &  J0347     &  H$\gamma$ $\lambda$ 4340 &  15.9         & H$\gamma$ $\lambda$ 4340  & 31.2         \\
                         &            &  H$\beta$ $\lambda$4861   &  24.3         & H$\beta$ $\lambda$4861 &  27.8         \\
                         & &                           & &                           &               \\
                         &  J1203     &  H$\gamma$ $\lambda$ 4340 &  12.4         &  H$\gamma$ $\lambda$ 4340 & --         \\
                         &            &  H$\beta$ $\lambda$4861   & 18.2          &  H$\beta$ $\lambda$4861 &  --        \\
                         & &                           & &                           &               \\
                         &  J0938     &  H$\gamma$ $\lambda$ 4340 &  24.7         &  H$\gamma$ $\lambda$ 4340 & --         \\
                         &            &  H$\beta$ $\lambda$4861   & 16.5          &  H$\beta$ $\lambda$4861 &  --         \\
                         & &                           & &                           &               \\
 Broad lines objects     &  J1158     &  H$\gamma$ $\lambda$ 4340 &  31.8         &  H$\gamma$ $\lambda$ 4340 & 32.2         \\
                         &            &  H$\beta$ $\lambda$4861   &  33.6         &  H$\beta$ $\lambda$4861 &  31.4         \\
                         & &                           & &                           &               \\
                         &  J0911     &  H$\gamma$ $\lambda$ 4340 &  19.8         &  H$\gamma$ $\lambda$ 4340 & 14.7        \\
                         &            &  H$\beta$ $\lambda$4861   &  11.1         &  H$\beta$ $\lambda$4861 &  12.3         \\
                         & &                           & &                           &               \\
                         &  J0802     &  H$\gamma$ $\lambda$ 4340 &  43.9         &  H$\gamma$ $\lambda$ 4340 & 60.5         \\
                         &            &  H$\beta$ $\lambda$4861   &  51.4         &  H$\beta$ $\lambda$4861 &  11.4         \\
                         & &                           & &                           &               \\
                         &  J0153     &  H$\beta$ $\lambda$4861   &  31.5         &  H$\beta$ $\lambda$4861 &  28.7        \\
                         & &                           & &                           &               \\
\hline
\label{narr_bro}
\end{tabular}
\end{center}
\end{table*}

\begin{figure*}
\includegraphics[width=1.0\columnwidth]{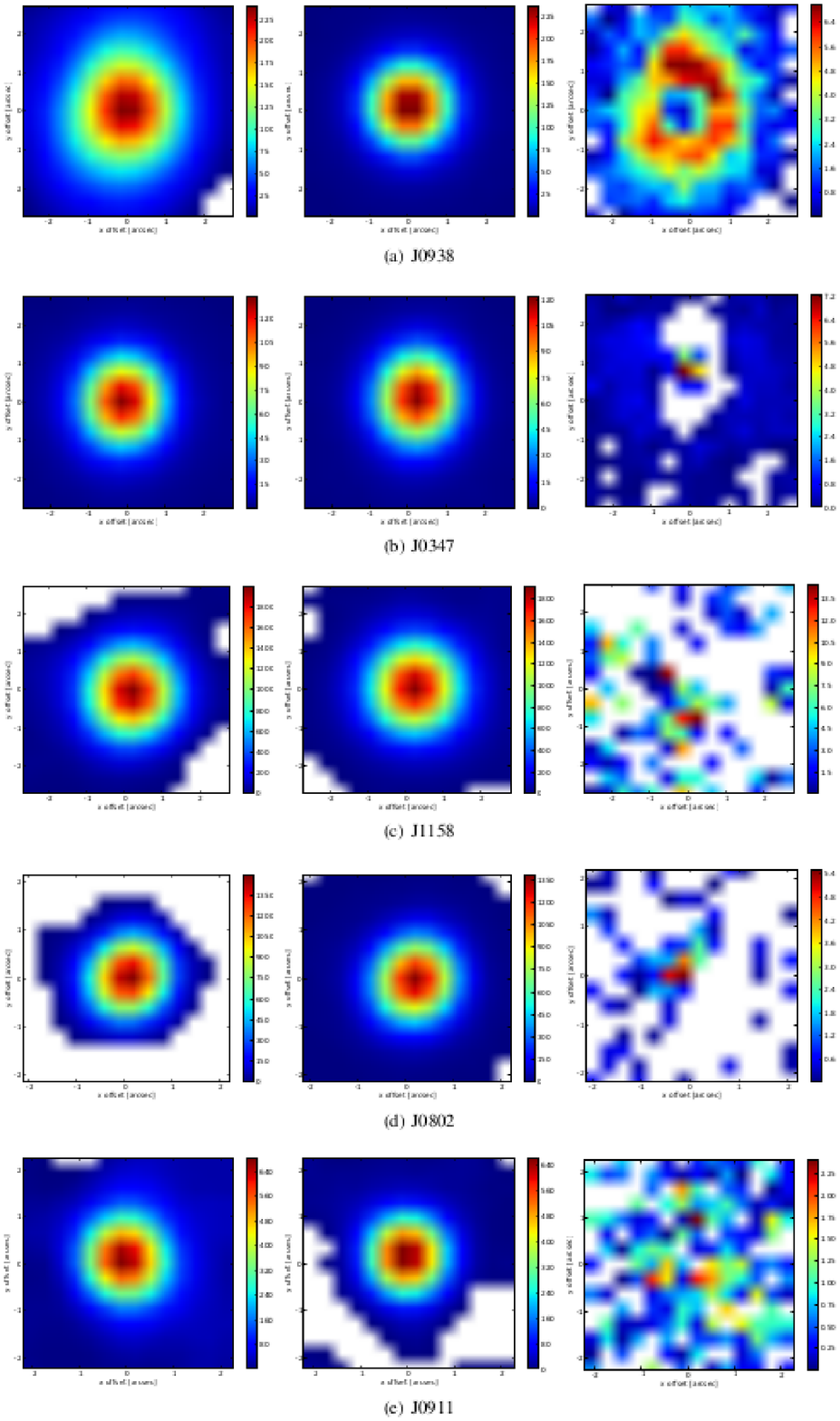}
\caption{The same as Fig.~\ref{fig:subtraction} but for the other galaxies.}
\label{fig:subtraction_rest}
\end{figure*}

\begin{figure*}
\includegraphics[width=0.9\linewidth]{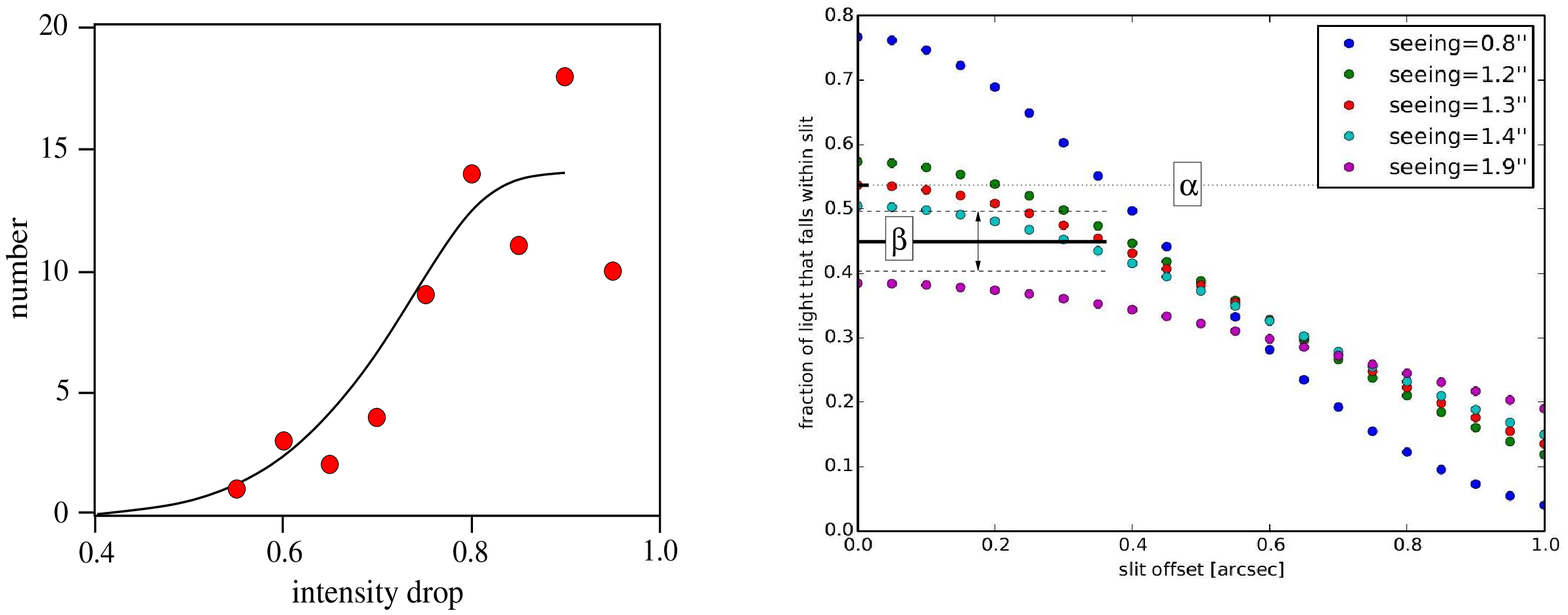}
\caption{
Left: For a 0.8 arcsec slit we show for all sources 
the distribution of the intensity drops of the 
fainter exposures with respect to the brightest once. The data is shown in 
bin widths of 0.05 (fat dots) and for comparison half the value obtained
with a bin width of  0.10 (black line).
Right: Intensity drop due to the combination of variable seeing and slit offset.
Case $\alpha$ indicates the level (dotted line) to which the spectra can be calibrated if 
for individual sources the fainter exposures are corrected to the level 
of the brightest once.
For case $\beta$ we assume that the brightest exposures
all are subjected to the mean drop (fat black line) indicated by the statistics of the 
faintest with respect to the brightest exposures.
The arrow and the dashed lines indicate the standard deviation from the mean.
}
\label{slitlosses}
\end{figure*}

\begin{figure*}
\begin{center}
\includegraphics[width=2.0\columnwidth]{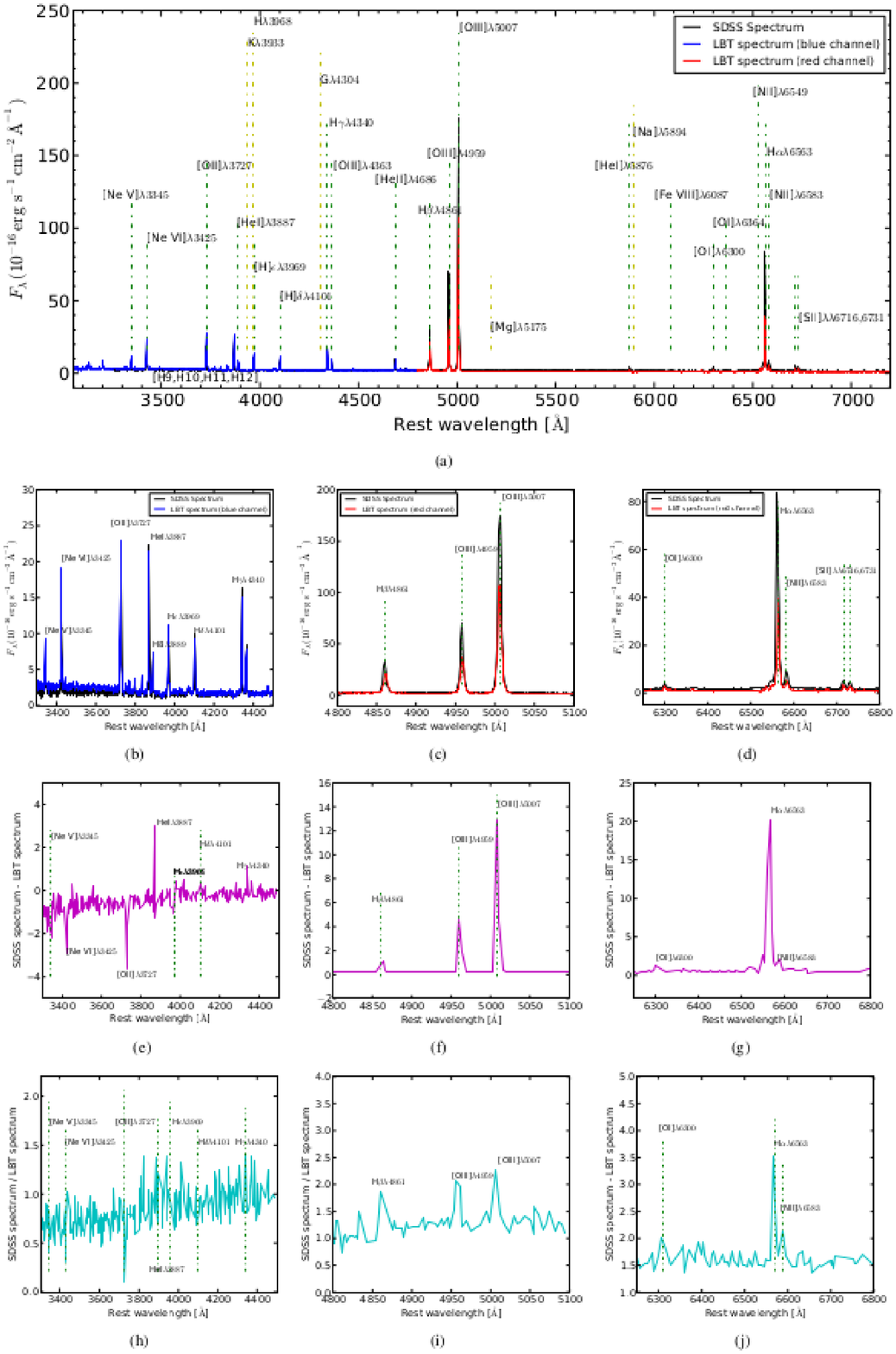}
\caption{
The optical spectrum of J1203 and the result of calculating the difference and ratio 
of spectra from different epochs.
}
\label{spec_j1203}
\end{center}
\end{figure*}

\begin{figure*}
\begin{center}
\includegraphics[width=2.0\columnwidth]{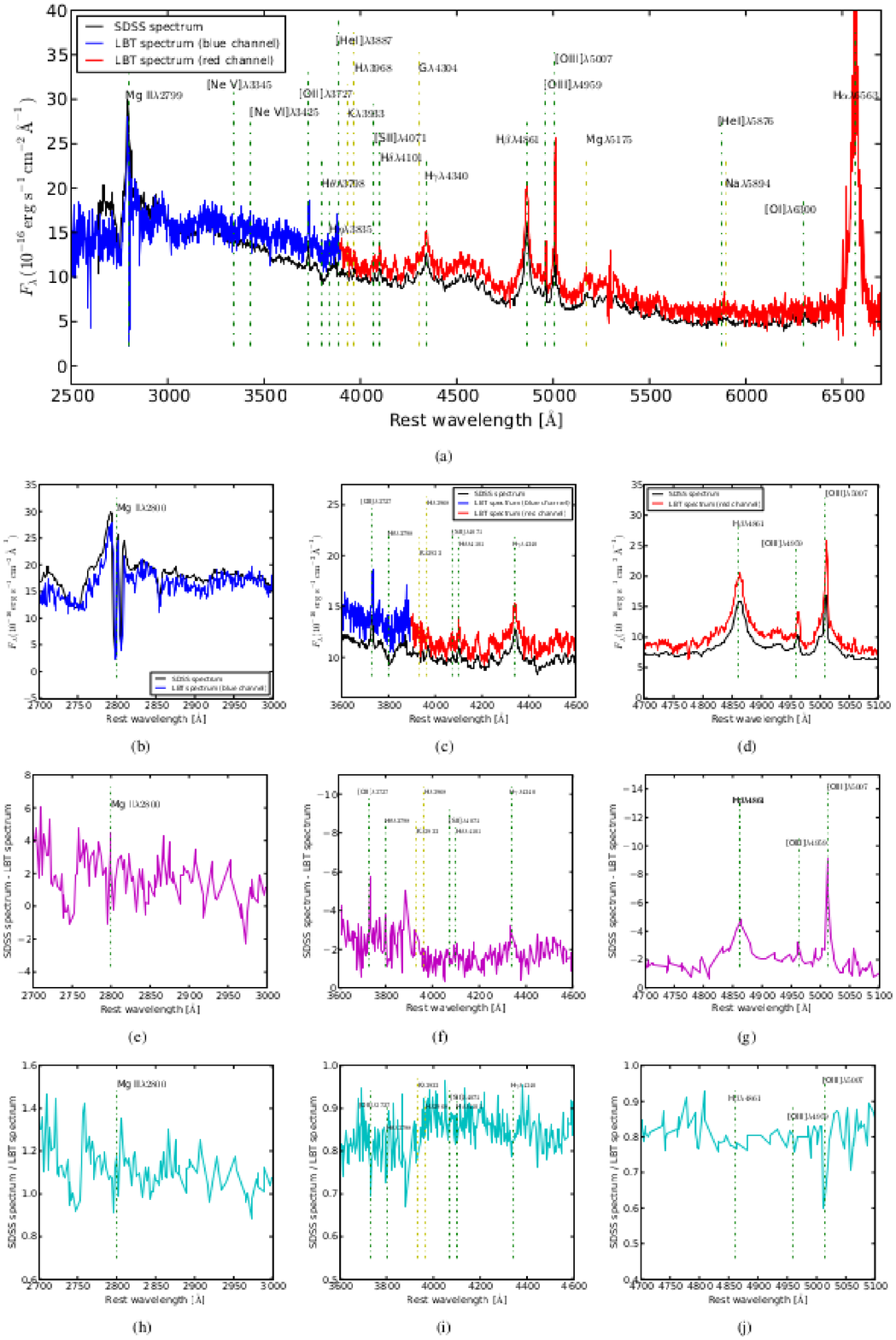}
\caption{
The optical spectrum of J1158 and the result of calculating the difference and ratio 
of spectra from different epochs.
}
\label{spec_j1158}
\end{center}
\end{figure*}

\begin{figure*}
\begin{center}
\includegraphics[width=2.0\columnwidth]{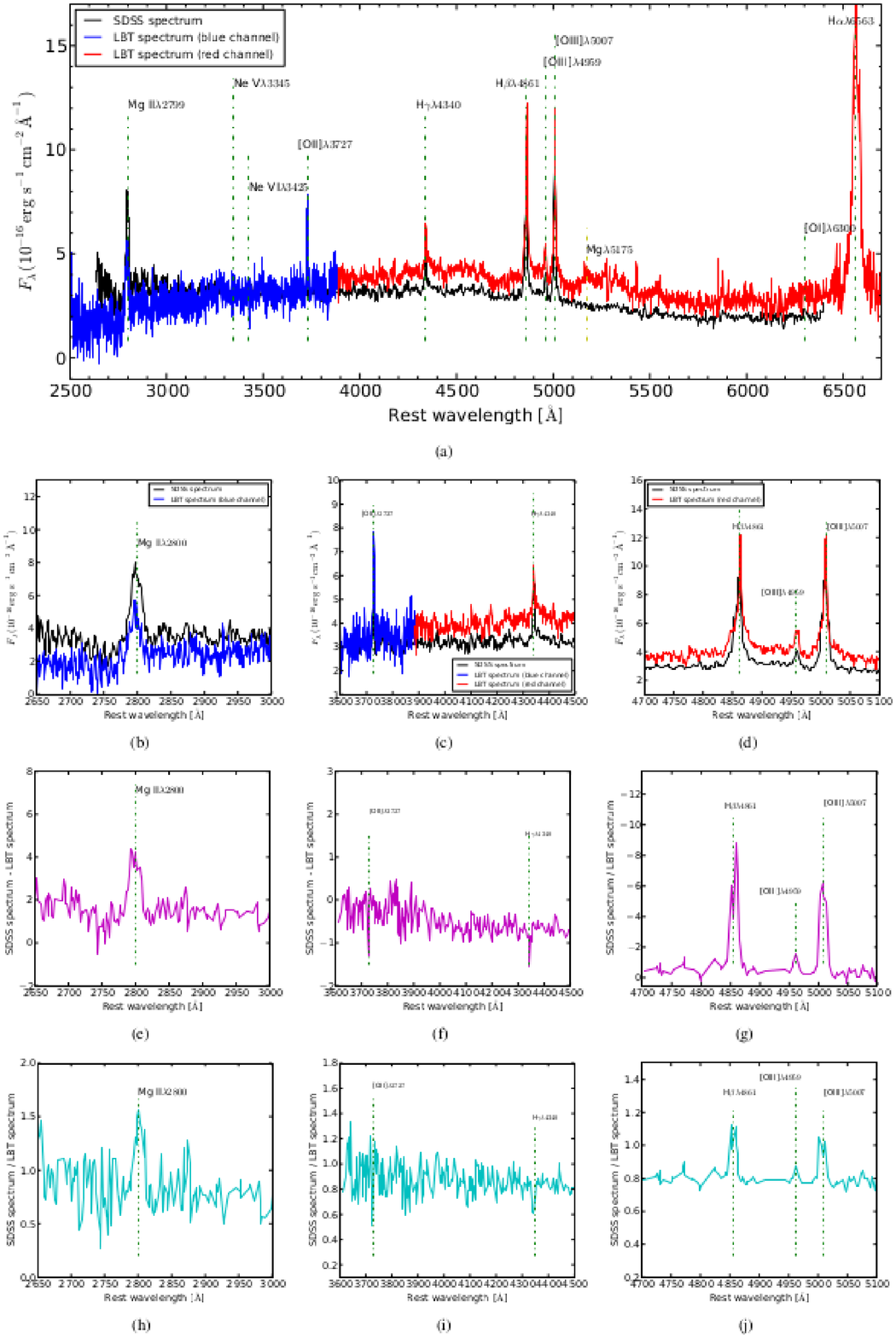}
\caption{
The optical spectrum of J0911 and the result of calculating the difference and ratio 
of spectra from different epochs.
}
\label{spec_j0911}
\end{center}
\end{figure*}

\begin{figure*}
\begin{center}
\includegraphics[width=2.0\columnwidth]{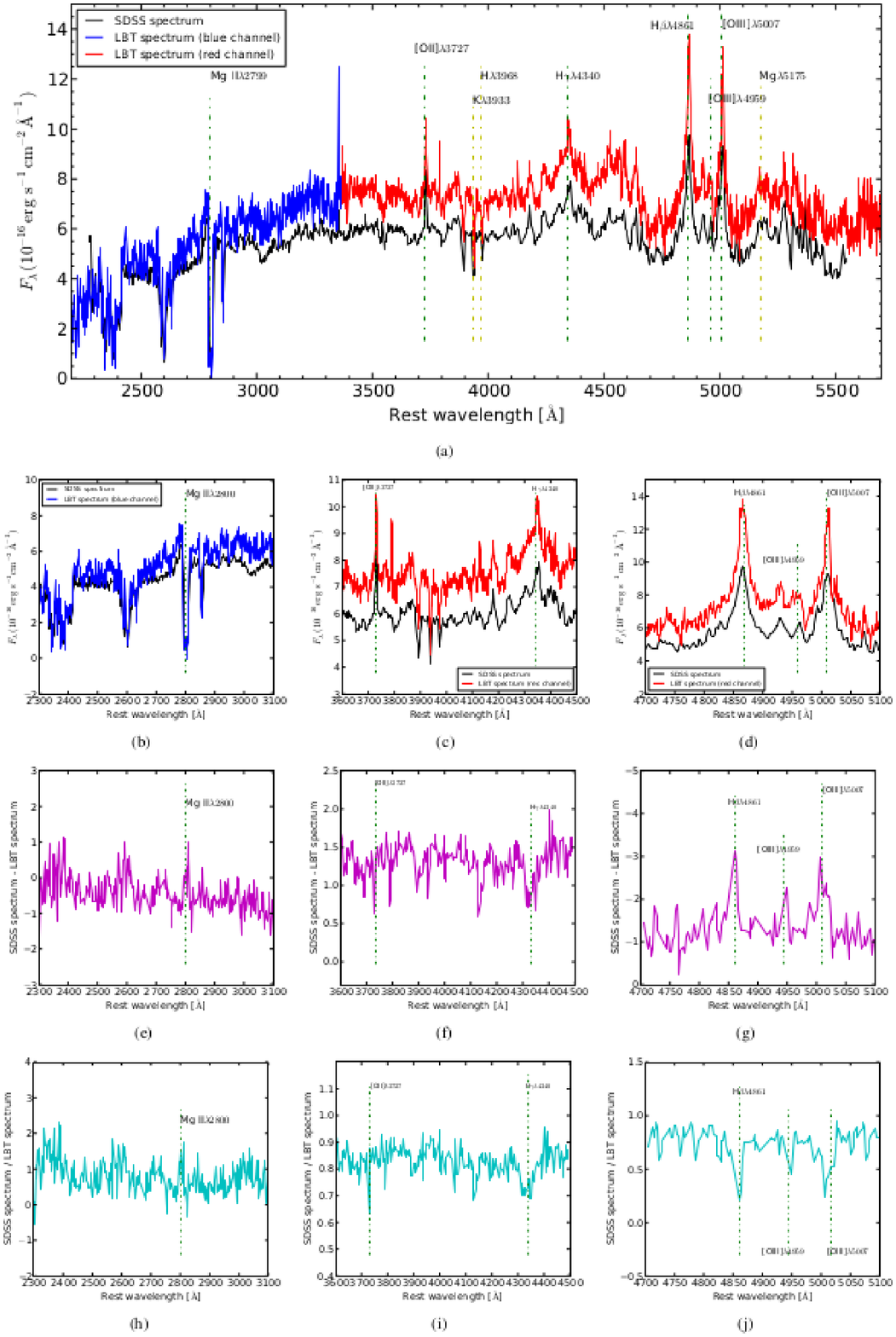}
\caption{
The optical spectrum of J0802 and the result of calculating the difference and ratio 
of spectra from different epochs.
}
\label{spec_j0802}
\end{center}
\end{figure*}

\begin{figure*}
\begin{center}
\includegraphics[width=2.0\columnwidth]{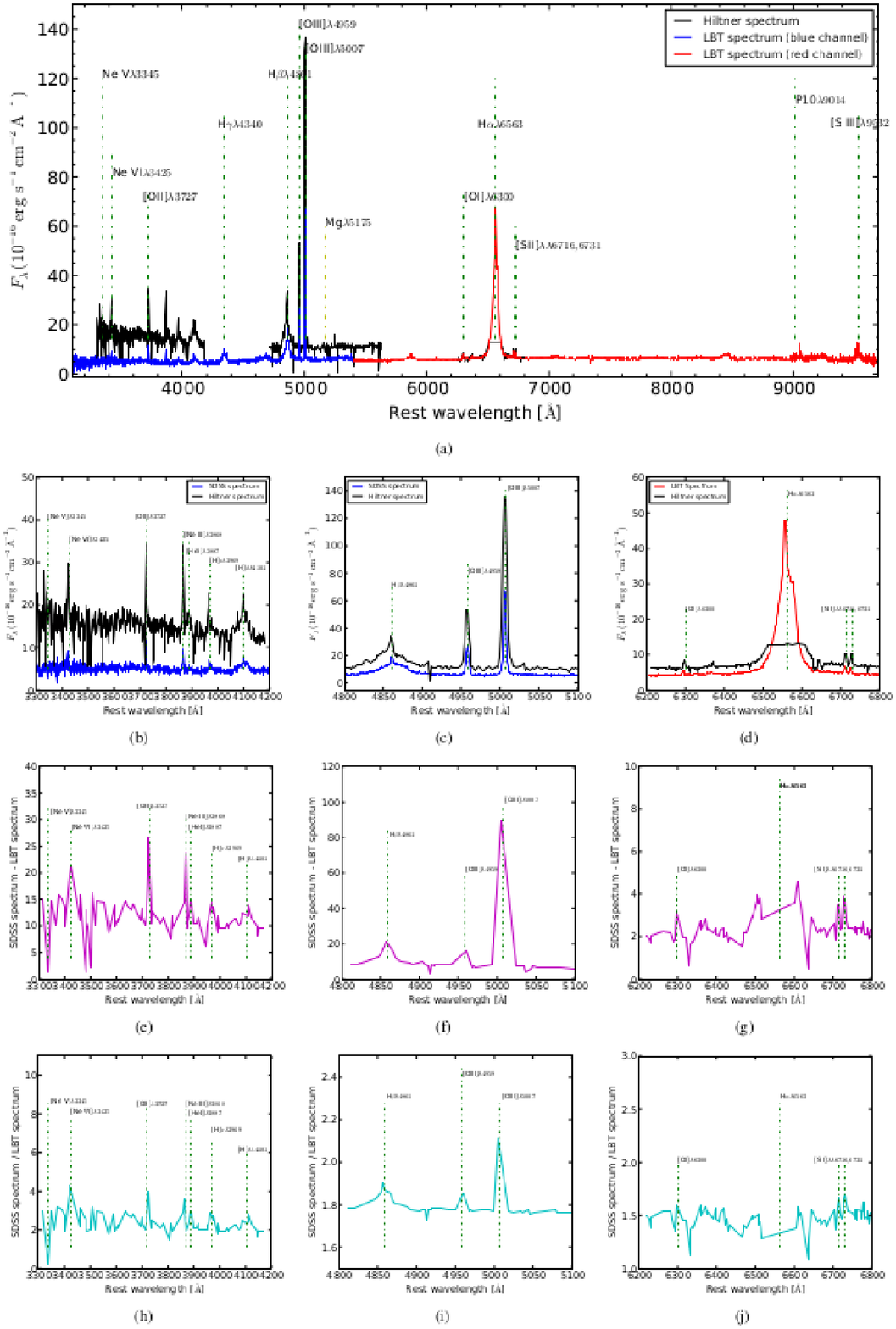}
\caption{
The optical spectrum of J0354 and the result of calculating the difference and ratio 
of spectra from different epochs.
}
\label{spec_j0354}
\end{center}
\end{figure*}

\begin{figure*}
\begin{center}
\includegraphics[width=2.0\columnwidth]{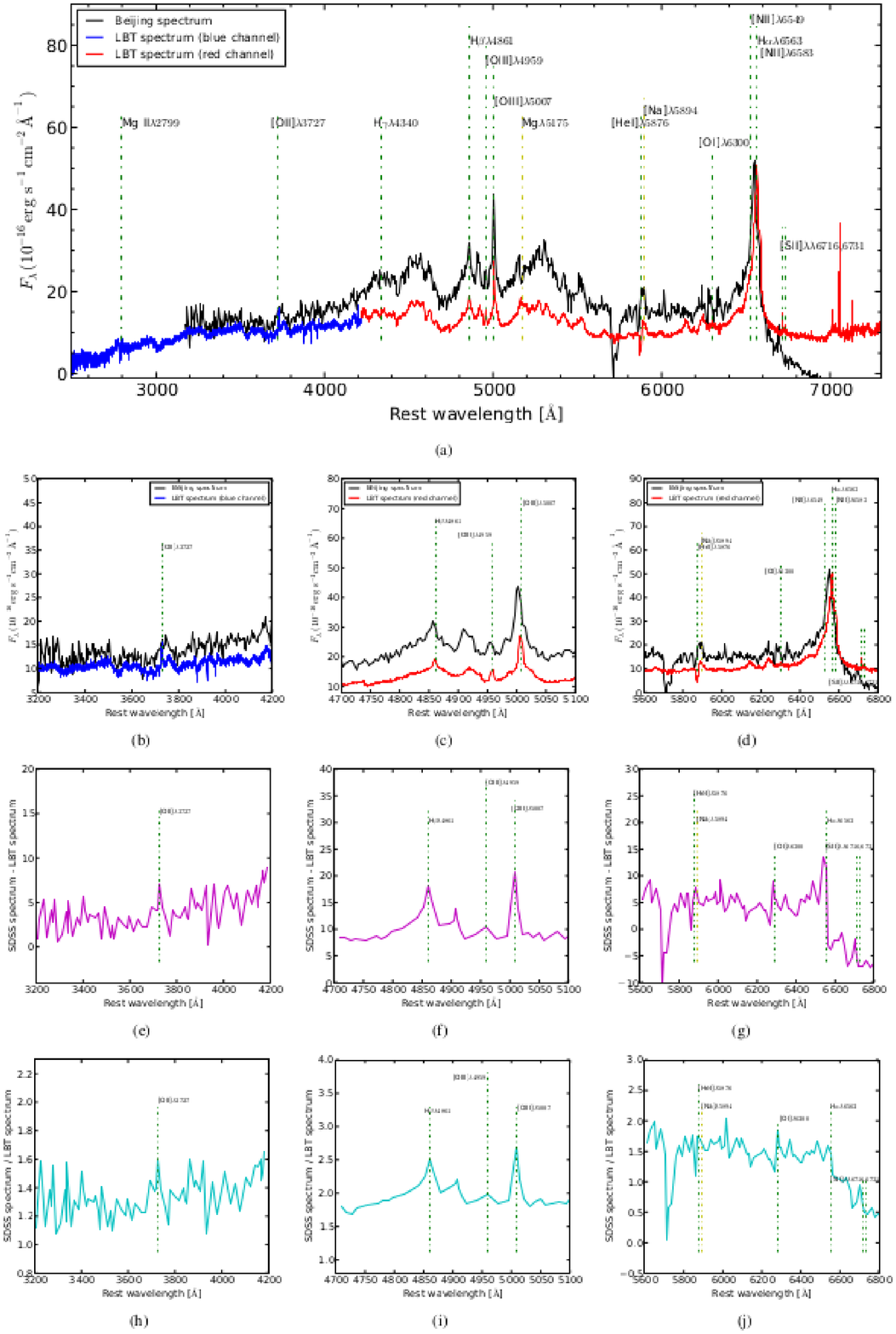}
\caption{
The optical spectrum of J0153 and the result of calculating the difference and ratio 
of spectra from different epochs.
}
\label{spec_j0153}
\end{center}
\end{figure*}

\begin{figure*}
\begin{center}
\includegraphics[width=2.0\columnwidth]{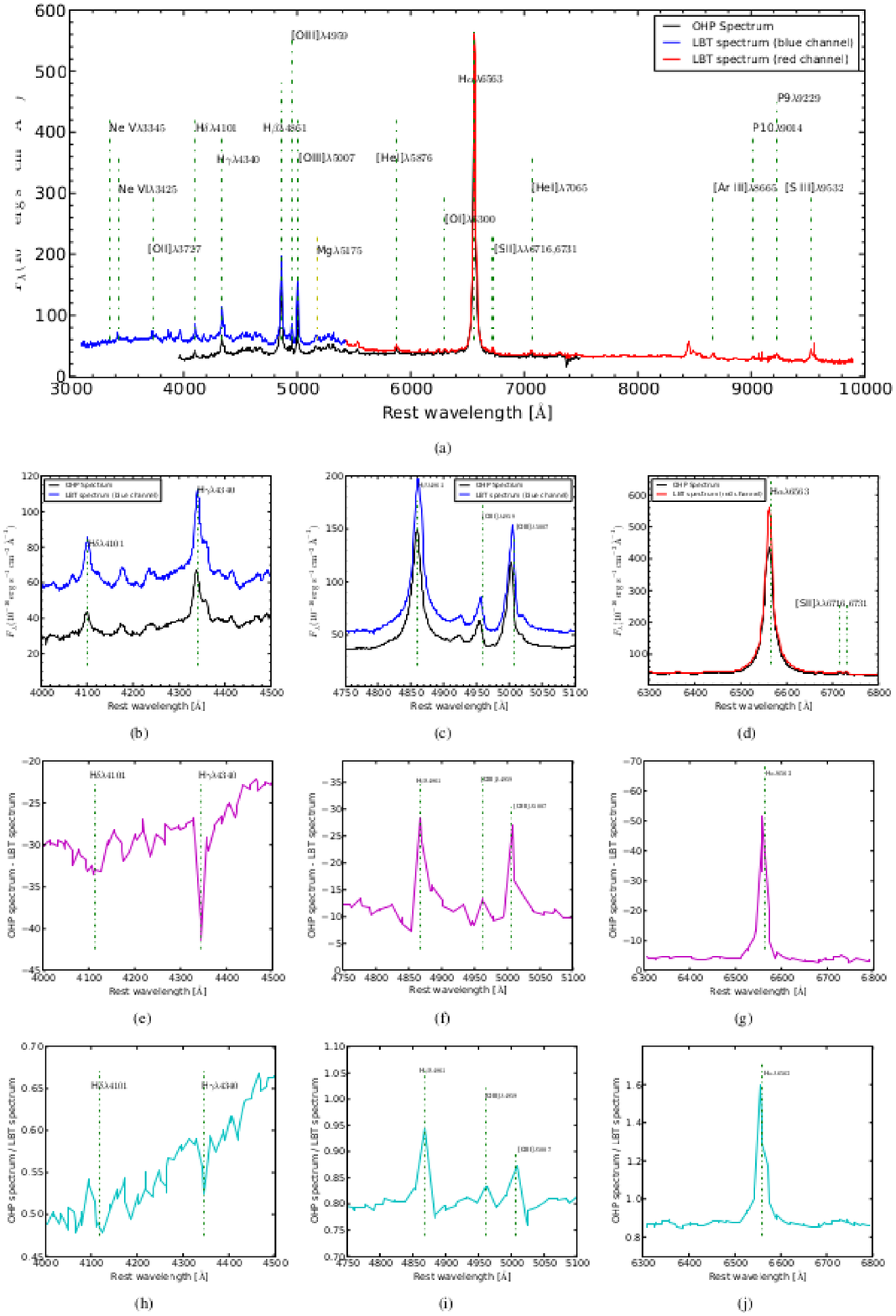}
\caption{
The optical spectrum of J0347 and the result of calculating the difference and ratio 
of spectra from different epochs.
}
\label{spec_j0347}
\end{center}
\end{figure*}

\begin{figure*}
\begin{center}
\includegraphics[width=0.5\linewidth]{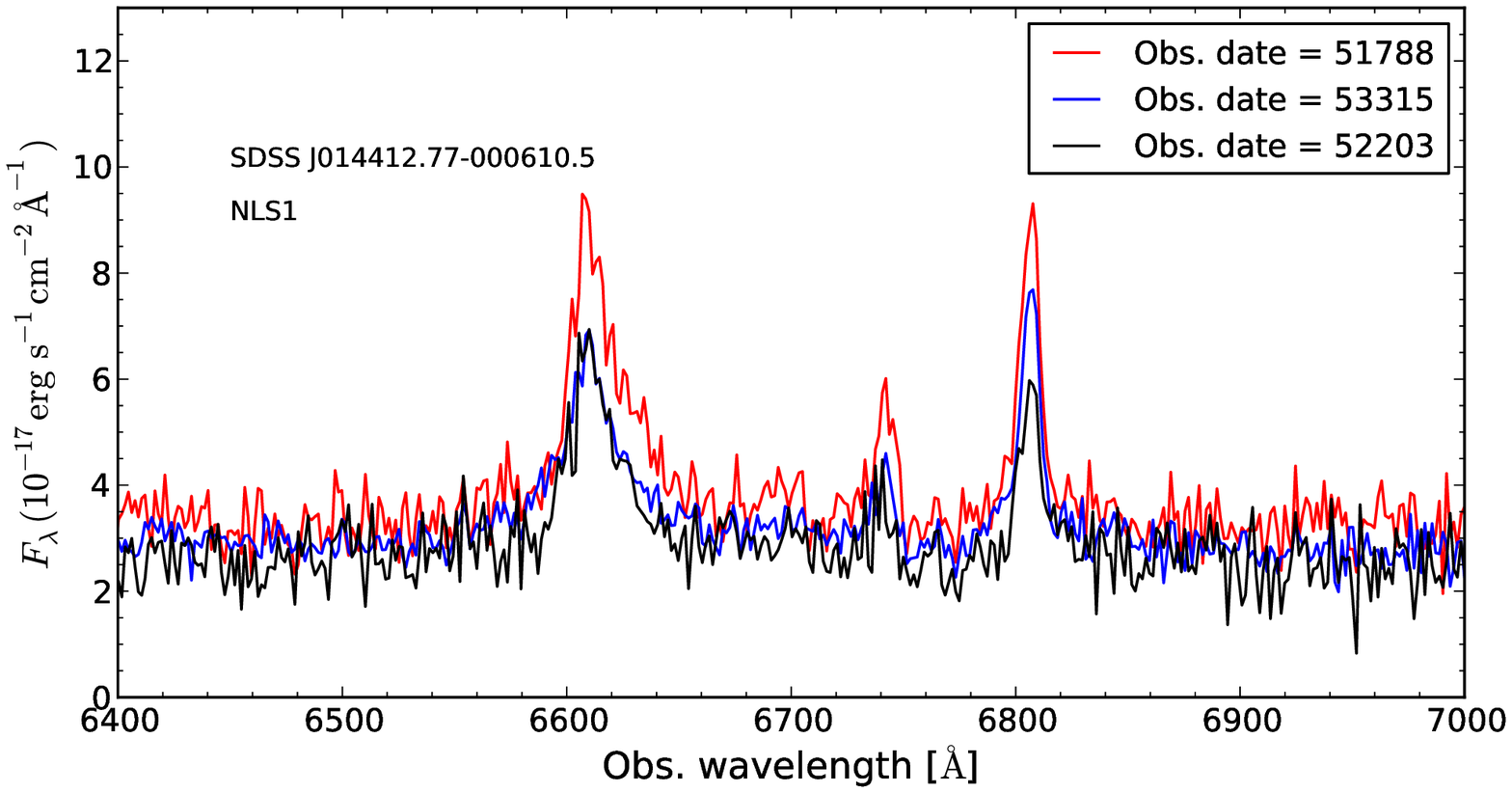}\\
\includegraphics[width=0.5\linewidth]{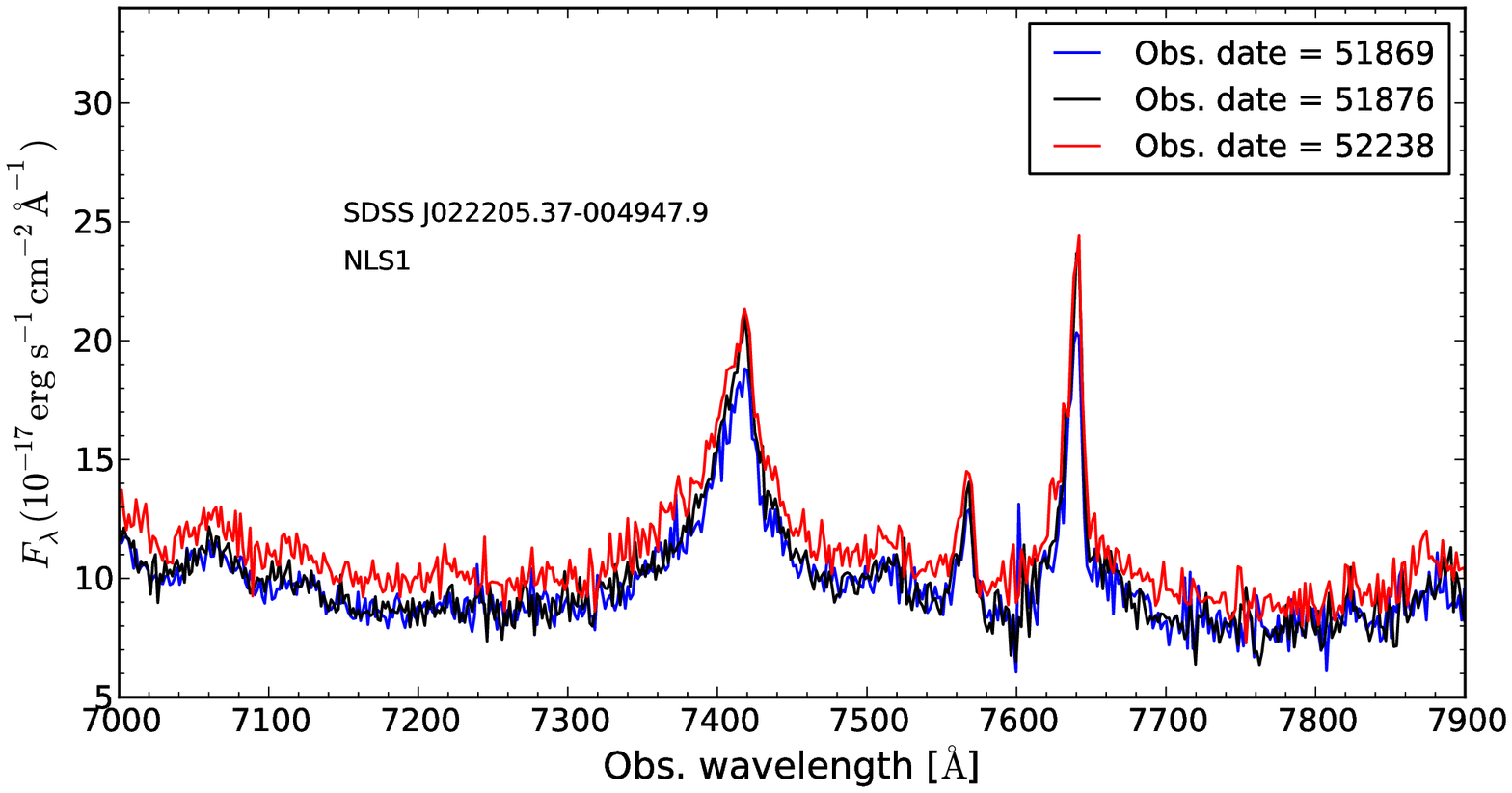}\\
\caption{
Three epoch SDSS spectra for J014412 and J022205.
These spectra allow us to derive variability estimates of 
$\Delta_{cont.}$=0.18 and $\Delta_{line.}$=0.45 for J014412 and 
$\Delta_{cont.}$=0.27 and $\Delta_{line.}$=0.26 for J022205.
These values lie well above the low calibration uncertainties for the SDSS data (section \ref{sec:SDSS}).
}
\label{SDSSnew}
\end{center}
\end{figure*}

\end{appendix}

\bsp

\label{lastpage}

\end{document}